\documentclass[preprint,12pt]{aastex}
\usepackage{emulateapj5}

\newcommand{\dd}{{\rm d}}
\newcommand{\sfrate}{{\cal S}}
\newcommand{\tin}{t_{\rm i}}
\newcommand{\tsf}{t_{\rm s}}
\newcommand{\zform}{z_{\rm form}}
\newcommand{\omstars}{\Omega_{\rm stars}}

\newcommand{\tdf}{2dF}
\newcommand{\grs}{2dFGRS}
\newcommand{\bj}{$b_{\rm J}$}
\newcommand{\cosmoparas}{$(h,\Omega_{m_0},\Omega_{\Lambda_0})$}
\newcommand{\sfr}{star-formation rate}
\newcommand{\sfh}{star-formation history}
\newcommand{\chisq}{$\chi^2$}
\newcommand{\sfrunits}{$h\,{\rm M}_{\odot}\,{\rm yr}^{-1}\,{\rm Mpc}^{-3}$}
\newcommand{\rhosfr}{$\rho_{\rm SFR}$}

\begin{document}

\title{The \tdf\ Galaxy Redshift Survey: 
  constraints on cosmic star-formation history from the cosmic spectrum}
\shorttitle{The 2dFGRS: constraints on cosmic \sfh}

\author{{} Ivan K.\ Baldry$^1$, Karl Glazebrook$^1$, 
Carlton M.\ Baugh$^2$, Joss Bland-Hawthorn$^3$, Terry Bridges$^3$, 
Russell Cannon$^3$, Shaun Cole$^2$, Matthew Colless$^4$, Chris Collins$^5$, 
Warrick Couch$^6$, Gavin Dalton$^7$, Roberto De Propris$^6$, 
Simon P.\ Driver$^8$, George Efstathiou$^9$, Richard S.\ Ellis$^{10}$, 
Carlos S.\ Frenk$^2$, Edward Hawkins$^{11}$, Carole Jackson$^4$, 
Ofer Lahav$^9$, Ian Lewis$^7$, Stuart Lumsden$^{12}$, Steve Maddox$^{11}$, 
Darren S.\ Madgwick$^9$, Peder Norberg$^2$, John A.\ Peacock$^{13}$, 
Bruce A.\ Peterson$^4$, Will Sutherland$^7$, Keith Taylor$^3$}
\shortauthors{I.\ K.\ Baldry et al.\ (the 2dFGRS collaboration)}

\affil{
$^1$Department of Physics \& Astronomy, Johns Hopkins University,
       Baltimore, MD~21218-2686, USA \\
$^2$Department of Physics, University of Durham, South Road, 
    Durham DH1~3LE, UK \\ 
$^3$Anglo-Australian Observatory, P.O.\ Box 296, Epping, NSW~2121,
    Australia\\
$^4$Research School of Astronomy \& Astrophysics, The Australian 
    National University, Weston Creek, ACT~2611, Australia \\
$^5$Astrophysics Research Institute, Liverpool John Moores University,  
    Twelve Quays House, Birkenhead L14~1LD, UK \\
$^6$Department of Astrophysics, University of New South Wales, Sydney, 
    NSW~2052, Australia \\
$^7$Department of Physics, University of Oxford, Keble Road, 
    Oxford OX1~3RH, UK \\
$^8$School of Physics and Astronomy, University of St Andrews, 
    North Haugh, St Andrews, Fife KY6~9SS, UK \\
$^9$Institute of Astronomy, University of Cambridge, Madingley Road,
    Cambridge CB3~0HA, UK \\
$^{10}$Department of Astronomy, California Institute of Technology, 
    Pasadena, CA~91125, USA \\
$^{11}$School of Physics \& Astronomy, University of Nottingham,
       Nottingham NG7~2RD, UK \\
$^{12}$Department of Physics, University of Leeds, Woodhouse Lane,
       Leeds LS2~9JT, UK \\
$^{13}$Institute for Astronomy, University of Edinburgh, Royal Observatory, 
       Blackford Hill, Edinburgh EH9~3HJ, UK \\
}

\begin{abstract}
  We present the first results on the history of star formation in the
  Universe based on the `cosmic spectrum', in particular, the
  volume-averaged, luminosity-weighted, stellar absorption line
  spectrum of present day galaxies from the 2dF Galaxy Redshift
  Survey. This method is novel in that unlike previous studies it is
  {\em not} an estimator based on total luminosity density. The cosmic
  spectrum is fitted with models of population synthesis, tracing the
  history of star formation prior to the epoch of the observed
  galaxies, using a method we have developed which decouples continuum
  and spectral-line variations and is robust against
  spectrophotometric uncertainties.  The cosmic spectrum can only be
  fitted with models incorporating chemical evolution and indicates
  there was a peak of \sfr\ in the past of at least three times the
  current value and that the increase back to $z=1$, assuming it
  scales as $(1+z)^\beta$, has a strong upper limit of $\beta<5$. We
  find in the general case there is some model degeneracy between star
  formation at low and high redshift. However, if we incorporate
  previous work on star formation at $z<1$ we can put strong {\em
    upper limits} on the \sfr\ at $z>1$: e.g., if $\beta>2$ then the
  SFR for $1<z<5$ scales as $(1+z)^\alpha$ with $\alpha<2$.  This is
  equivalent to stating that no more than 80\% of stars in the
  Universe formed at $z>1$.  Our results are consistent with the
  best-fit results from compilations of cosmic SFR estimates based on
  UV luminosity density, which give $1.8 < \beta < 2.9$ and $-1.0 <
  \alpha < 0.7$, and are also consistent with estimates of $\omstars$
  based on the $K$-band luminosity density.
\end{abstract}


\section{Introduction}

The analysis of the comoving \sfr\ (SFR) density as a function of
redshift has been the subject of much recent work.  The onset of large
redshift surveys at $z<1$ \citep[e.g.][]{LLHC96} and $z>3$
\citep[e.g.][]{steidel99} has allowed the volumetric emission of
luminosity in different bands to be traced with redshift. In
particular from these studies there is now good evidence for a rise in
\sfr\ by a factor of about 8 between $z=0$ and $z=1$ \citep{Hogg02}
and either a $z>1$ decline \citep{madau96} or plateau
\citep{pettini98}.

Most measurements of the SFR to date have been based on some type of
{\em luminosity density} which is thought on theoretical and/or
empirical grounds to trace the \sfr. This use of luminosity per unit
volume reflects an attempt to decouple the stellar history of the
Universe from its dynamical history. Other statistical measures such
as object counts versus luminosity or redshift are affected by the
changes in the number of galactic objects by merging as well as
evolution of stellar populations. In contrast, the change in the light
budget per unit volume is only affected by the stellar production in
that volume regardless of the changes in the number of objects. Thus,
for example, it makes sense to compare the total production of stars
with the total metal abundance today \citep{CLGM88}.

Popular tracers of \sfr\ include: the UV 1500--3000\AA\ continuum
(\citealp{conn97}; \citealp*{MPD98}); the radio continuum
\citep{MCGH99}; emission in the H$\alpha$ and H$\beta$ lines
\citep{glaze99}, and; other line emission such as Ly$\alpha$
\citep{kudritz00} or O{\small II} \citep{CETH90,HCBP98}.  The far-IR
thermal dust emission has also been used to trace \sfr\ 
\citep{hughes98} although so far without the benefit of redshift
information. All these measures have in common the use of some kind of
luminosity per unit volume whose change is proportional to the \sfr\ 
(with corrections).  The debate over the $z>1$ slope reflects the
uncertainty in the dust correction to the UV continuum measurements,
which are the easiest to measure at high redshift but most affected by
dust. The other indicators are harder to measure and are affected by
small number statistics.

In this paper, we present new constraints on the history of star
formation based on the ensemble stellar populations of relatively
nearby present day galaxies ($z<0.3$). The concept is to use the
average spectrum of nearby galaxies to constrain the earlier history
of star formation leading up to that stellar population. The average
spectrum contains absorption features for stars of all ages and probes
look-back times of 0.2--10 Gyr. With the advent of the large galaxy
redshift surveys, such as the Two-degree Field Galaxy Redshift Survey
(\grs) and the Sloan Digital Sky Survey main galaxy sample (SDSS-mgs),
it has become feasible to combine the spectra of $10^4$--$10^5$
galaxies to form very high signal-to-noise intermediate resolution
spectra that represent the average emission of the universe at various
redshifts ($0.03 \la z \la 0.25$).  Effectively, the surveys can be
regarded as having a series of apertures on the cosmic background
emission rather than apertures on individual galaxies.

What is novel about this method is that it is not based on any
luminosity output with time of the Universe, unlike all the other
indicators discussed above. The method uses an integral over the \sfh\
rather than attempting to track the derivative (the SFR) and uses the
whole visible spectrum at intermediate resolution.  Perhaps the
nearest approach to this in the past has been the work of
\citet{abrah99mar} who used the color distribution of $z\sim0.5$
galaxies to derive their star-formation histories which were then
combined to form a cosmic \sfh. Recently, \citet*{HIC01} estimated the
global SFR density from \sfh\ measurements of the Local Group. The
problem here is that the Local Group SFR may not represent the cosmic
mean and is subject to large cosmic variations such as recent
``mini-bursts'' of star formation in the Milky Way.

An advantage of this kind of `fossil cosmology' approach over the
direct measurement is a reduced sensitivity to extinction. Young
stars are born in dusty regions, this plagues the direct measurement
approach. When they age they migrate out of such regions and
contribute to the older stellar populations we observe.

In this paper, we describe the application of this method to 166\,000
spectra in the redshift range 0.03--0.25 from the \grs\ \citep{colless01}.
The plan of this paper is as follows. In Section~\ref{sec:data} we
describe the \grs\ data and our methods for combining the spectra. In
Section~\ref{sec:scenarios} we describe our analytic models for cosmic
star-formation scenarios. In Section~\ref{sec:results} we describe our
fitting procedure and the best-fit models. In Section~\ref{sec:biases}
we discuss the impact of possible biases on our results from aperture
effects and luminosity selection. Finally in Section~\ref{sec:conc} we
give our conclusions.

\section{The \grs\ data} \label{sec:data}

The 2dF Galaxy Redshift Survey is a magnitude-limited spectroscopic
survey \citep{maddox98,colless99,colless01} using the Anglo-Australian
Observatory's 2dF facility which is capable of observing up to 400
galaxies simultaneously \citep*{TG90,LGT98,lewis02}.  The magnitude
limit of the survey is an extinction-corrected \bj\ of 19.45 selected
from the Automated Plate Measuring (APM) galaxy catalogue
(\citealp*{MES90,MES96}; \citealp{MESL90}). By the end of the \grs\ 
survey (2002 January), up to 250\,000 unique galaxy redshifts are
expected to have been measured.  The survey covers approximately 2000
deg$^2$ of sky distributed between the NGP and SGP in high-galactic
latitude fields.  A full description of the survey geometry is given
by \citet{colless01}.

In 2001 June, about 173\,000 unique galaxy redshifts had been
measured.  This is the sample used in the analysis presented in this
paper.  The spectra are observed through a fixed $2''\!.1$-fiber
aperture and the wavelength coverage varies only slightly from
observing run to run, consistently covering the range 3700--7860\AA\ 
with a 2.1-pixel resolution FWHM of 9.0\AA.  All the spectra in the
survey have been eyeballed and assigned a quality $Q$ from 1 to 5
\citep{colless01}: 1, no identifiable redshift; 2, a possible redshift; 3,
a 90\% reliable redshift; 4, a 99\% reliable redshift; 5, a 99\%
reliable redshift with a high-quality spectrum.  The survey is
considered to consist of those galaxy spectra with $Q\ge3$
(approximately 92\% of galaxies observed).  These galaxies have a
median redshift of about 0.11 and the typical spectral signal-to-noise
ratio at the survey limit is about 10 per pixel.

We construct our `cosmic spectra' in redshift slices $z \rightarrow z
+ \Delta z$ by applying an instrument response correction,
de-redshifting to the rest frame and summing up all the $Q\ge3$
spectra in the interval $\Delta z$. The galaxies are scaled to match
their measured \bj\ luminosity by comparison with an integration of
the \bj\ filter curve over the measured spectrum (in the observed
frame).  The scaling allows for the fact that the fibres sample only a
fraction of a galaxy's light and for extinction and exposure-time
variations between observations. A maximum scaling is allowed (as a
function of apparent magnitude) to avoid adding excessive noise from
poor-quality spectra. This scaling limit is only applied to 5\% of the
data. Finally the cosmic spectrum is normalized to a mean of unity
over a set wavelength range -- to re-iterate, our method of analysis
uses spectral features not absolute luminosities other than for the
weighting of the galaxy spectra.  This spectrum represents the
spectral emission per unit volume in the $z \rightarrow z + \Delta z$
interval down to the limiting magnitude of the survey.  As a result of
the \bj$\approx 19.45$ limit of the \grs, the corresponding limiting
absolute luminosity will be higher at higher redshifts.  This
exclusion of lower luminosity galaxies is a possible source of bias.
This is quantified in Section~\ref{sec:biases}.

The instrument response correction consists of applying the average
2dF spectral response as given by \citet{lewis02}. Only a simple
scaling is applied for light lost outside the fixed angular size fiber
aperture which again is another potential source of bias to be
discussed in Section~\ref{sec:biases}.

\begin{figure*}
\centerline{
\epsfxsize=18.4cm
\epsfbox{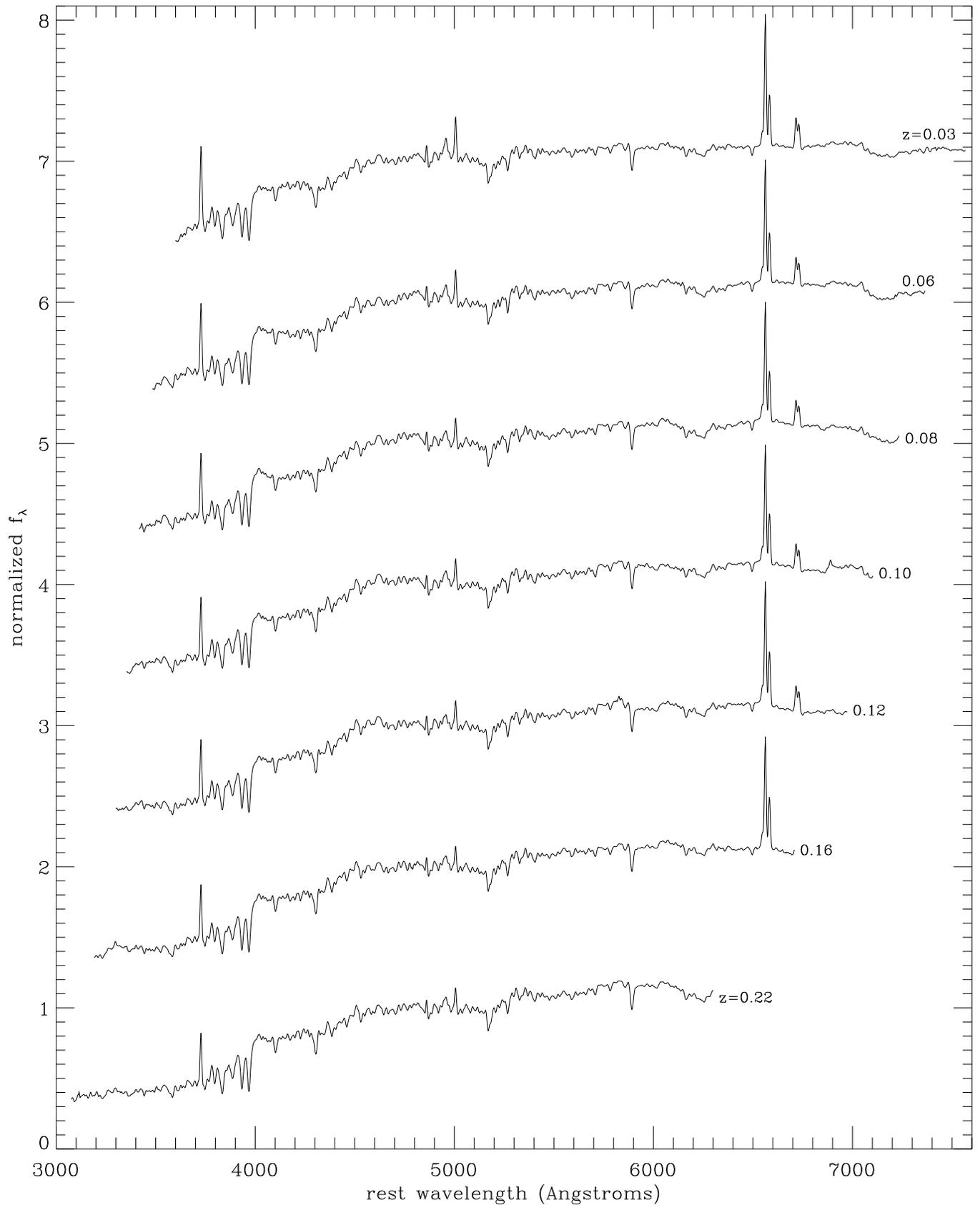}
}
\caption{Averaged \grs\ spectra from various redshift bins{}
  (0.025--0.04, 0.06--0.07, 0.08--0.09, 0.10--0.11, 0.12--0.13,
  0.155--0.175, 0.20-0.25).  The spectra were normalized to a mean of
  1.0 between 4200 and 5800\AA, and offset by 1.0 from each other.}
\label{fig:averaged-spectra}
\end{figure*}

Examples of cosmic spectra, at a series of redshifts, are shown in
Figure~\ref{fig:averaged-spectra}. As might be expected from a broad
\sfh, the average emission from the Universe looks remarkably like an
Sb--Sbc galaxy spectrum \citep{Kenn92}.

\section{Cosmic star-formation scenarios} \label{sec:scenarios}

First we must decide how best to parameterize cosmic star-formation
histories. We use two different families of parameterization.

The first is a physically motivated parameterization, using an infall
model to describe star formation as a function of time and timescales.

The second is a purely empirical model, which describes star formation
as a series of power laws with redshift. This facilitates direct
comparisons with the literature.

For the cosmology we consider two types, both constrained to be flat
in accord with the Cosmic Microwave Background data \citep{bernard00}:

\begin{enumerate}
\item The emerging standard cosmology (which we denote $\cal C$1),
  which has cosmological parameters $C_0 =$ \cosmoparas\ $=$
  (0.70,0.3,0.7)\footnote{$h=H_0{ / 100\,{\rm km\,s^{-1}\,Mpc^{-1}}}$}.
  These parameters are the current best constraints from a variety of
  observations \citep{Silk99}.
\item A longer-age cosmology (which we denote $\cal C$2) with $C_0 =$
  (0.55,0.2,0.8).  These parameters have been adjusted on the longer
  age side to the margins of consistency with modern limits.
\end{enumerate}

For galaxies at $z=0.1$ with $\zform=5$, these cosmologies give ages
of 11.0 and 15.7\,Gyr, respectively.

Given a \sfh\ and a cosmology we can compute spectra using standard
evolutionary synthesis codes. We used the PEGASE code
\citep{FR97}\footnote{PEGASE Version 2, revised 2001 May 5, URL {\tt
    http://www.iap.fr/users/fioc/PEGASE.html}}.  Both
parameterizations require an initial-mass function (IMF) and we choose
the \citet{Salp55} power law slope with stellar mass in the range
0.1--120\,M$_{\odot}$. We have also investigated the \citet{Kenn83} IMF
over the same mass range. We also have choices to make on metallicity
evolution and dust extinction. For the former, the enrichment of the
inter-stellar medium is determined using the calculations of
\citet{WW95} within the PEGASE code. For the dust extinction, the
prescription for an inclination-averaged disk geometry is used.
However, the choice of extinction, E$(B-V)\sim0.2\pm0.1$, makes
negligible difference to the results in this paper for two reasons:
(a) we are only determining the relative SFR as a function of time or
redshift, i.e., not comparing predicted luminosity with luminosity
density, and; (b) the robust results are primarily constrained using
high-pass filtered spectra which are insensitive to the extinction
model. We note that there will be a second-order effect from any
systematic variation of extinction with spectral type and/or
luminosity. 

\subsection{Physical parameterization} \label{sec:phys-param}

\begin{figure*}[ht]
\centerline{
\epsfxsize=9.2cm
\epsfbox{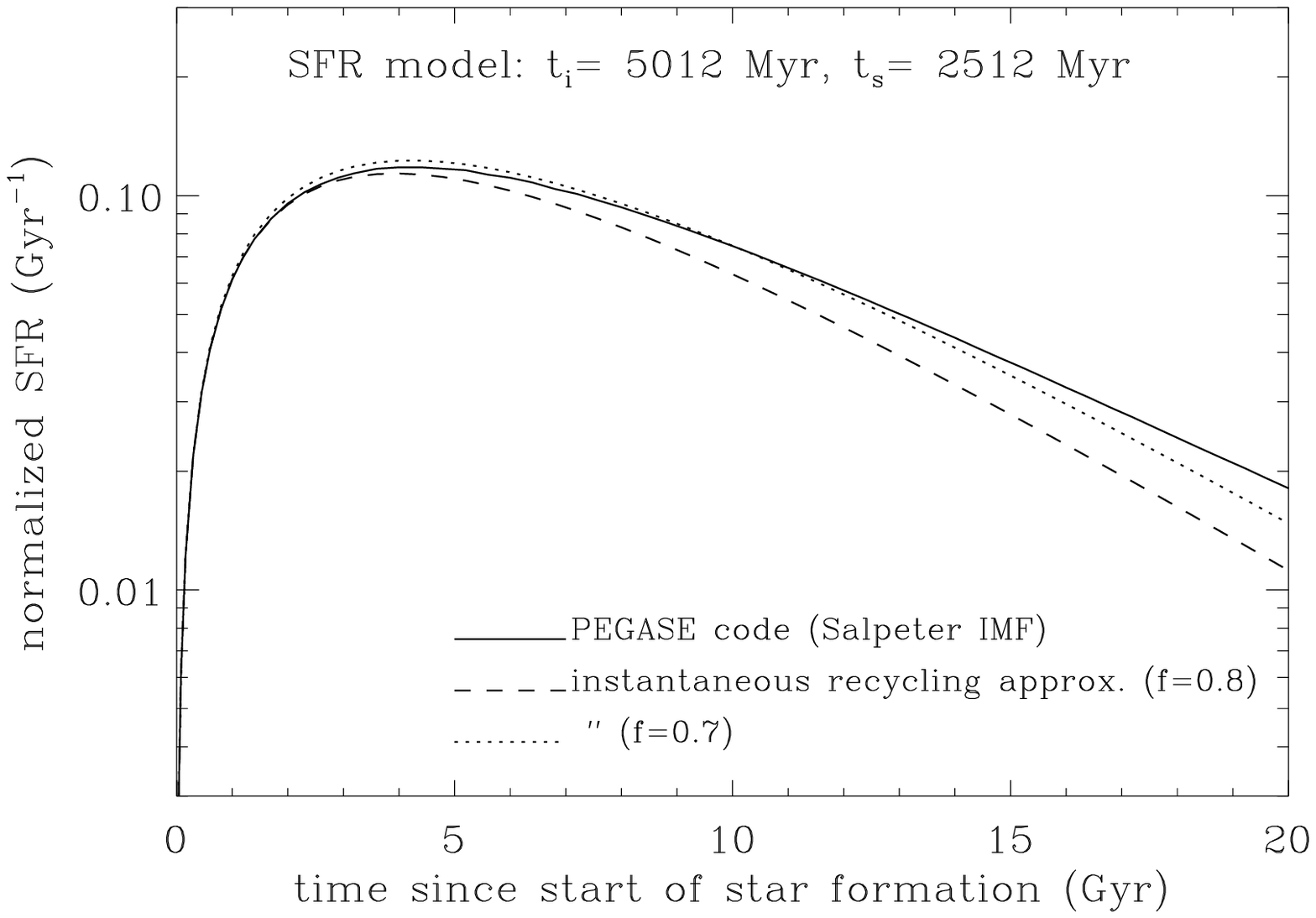}
}
\caption{Example star-formation scenario{}
  using an infall timescale and a star-formation timescale. The
  instantaneous recycling approximation, Equation~\ref{eqn:sfr-model},
  assumes a $R$ ($= 1-f$) mass-fraction of stars formed are returned
  instantaneously to the inter-stellar medium. The normalized SFR is
  relative to the total mass of gas available for forming stars, the
  integral of the SFR can be greater than unity because of recycling of
  material back to the ISM.}
\label{fig:example-scenario}
\end{figure*}

The first parameterization is a natural scenario. Star formation
starts at $z=\zform$. Gas falls into sufficiently dense regions to
form stars with an infall timescale $\tin$:
\begin{equation}
M_{\rm galaxy} = 1 - e^{-t/\tin} 
\end{equation}
The gas in these regions forms stars at a rate ($\sfrate$) proportional
to the amount of gas available with a star-formation timescale $\tsf$:
\begin{equation}
\sfrate\, = {M_{\rm gas}}/{\tsf}
\end{equation}
This parameterization is readily implemented in the PEGASE population
synthesis code \citep{FR97}.  The code includes recycling of ejecta
into the inter-stellar medium (ISM) and consistent evolution of the
metallicity.  An analytical approximation to the \sfr\ can be
determined assuming an instantaneous-recycling approximation ($f$ is
the mass fraction of stars that are {\em not} returned to the ISM).
Using this prescription the \sfr\ evolves as
\begin{equation}
\frac{\dd\sfrate}{\dd t} = \frac{1}{\tsf}
\left( - f \sfrate + \frac{e^{-t/\tin}}{\tin} \right) \: ,
\end{equation}
For $f \tin \ne \tsf$, the solution for the \sfr\ with time is given by
\begin{equation}
\sfrate = \frac{ e^{-t/\tin} -  e^{-ft/\tsf} }{f\tin - \tsf}
\label{eqn:sfr-model}
\end{equation}
with initial condition $\sfrate = 0$ at $t = 0$.  The normalization is
such that the total mass of gas available is unity. This
parameterization is useful in that it is physically motivated and that
it allows for a rise at early times and a fall at late times of the
universal \sfr.  The deviation between the approximation and the code
becomes apparent when recycling from lower mass stars becomes
significant.  An example is shown in Figure~\ref{fig:example-scenario}
with $f=0.7$ and 0.8 for the recycling approximation.

\subsection{Empirical parameterization} \label{sec:emp-param}

In the second parameterization we empirically make the \sfr\ a
function of redshift.  In particular, we make the \sfr\ between $z=5$
(our choice of formation redshift) and $z=1$ proportional to
$(1+z)^{\alpha}$ and between $z=1$ and $z=0$ proportional to
$(1+z)^{\beta}$, with matched SFR at $z=1$. This form allows us to
make a comparison of the best-fitting parameters with previous studies
which measure \sfr\ with redshift via luminosity densities. In
particular we can compare the value of $\beta$ defined and measured by
\citet{Hogg02} as well as other measurements of the variation of the
comoving SFR density with redshift \citep[e.g.][]{LLHC96,madau96}.

The total amount of star formation ($r$) between $z=5$ and $z=0$ as a
fraction of the mass available was normalized to a range of values
from $r=0.3$ to $r=1.4$ (the SFR normalization).  In other words, the
mass of gas available for star formation is unity and the $r$ is the
total mass of stars formed since $\zform$. This can be greater than
unity because of recycling of material back into the inter-stellar
medium.  Higher $r$ values result in higher average metallicity
because the fusion products released into the ISM by supernovae are
more abundant relative to the remaining gas.

\section{Best-fit star-formation scenarios} \label{sec:results}

\subsection{Reductions}

The \grs\ data were divided into 14 redshift bins between $z=0.025$
and $z=0.25$, each containing about 12\,000 spectra.  Half the
redshift bins were below $z=0.11$ (the median redshift of the survey).
The lower and higher groups of 7 redshift bins were considered
separately and together to test the robustness to varying aperture and
selection effects (see Section~\ref{sec:biases} for a further
discussion of this).

For each redshift bin, the spectra were divided into ten positional
bins based on their coordinates (4 regions in the NGP and 6 in the
SGP).  A normalized average spectrum was calculated and the positional
bins were used to estimate the uncertainties. This spectrum represents
the total optical emission of all galaxies in the volume of the
redshift shell down to the \grs\ magnitude limit.  The absolute
magnitude limit is fainter than or about $M^*$ out to a redshift of
0.2 (the depth is discussed further in Section~\ref{sec:biases}).

To flux-calibrate the spectra they were divided by the 2dF response
function of \citet{lewis02} before coadding the fluxes at their rest
wavelengths.  Spectra contribute to the final spectrum in proportion
to their \bj\ luminosity. In addition, the averaged spectra were
smoothed and resampled to match approximately the much lower 20\AA\ 
resolution of the spectral library \citep*{LCB97} used by the
population-synthesis code.

\subsection{Goodness of fit} \label{sec:fit}

To evaluate the `goodness of fit' between a star-formation scenario
and the \grs\ data, we compare the spectrum from each redshift bin
with the appropriate model spectrum (at the same age) from the scenario.
Fiber spectra are known to be difficult to flux accurately so it is
desirable to develop methods which are insensitive to small 
spectrophotometric uncertainties.

The first method we used was to allow for the possibility of
spectrophotometric calibration errors by including a correction
function before evaluating the fit. This was to account for
spectrophotometric discrepancies between the 2dF response curve used
to calculate the average spectra and the true average \grs\ response.
We used a fourth order polynomial with {\em observed} wavelength for
this spectrophotometric correction, the coefficients being determined
by ratioing the model and data spectra (excluding the strongest
emission lines as per the fitting).  An important point is that the
correction is constrained to be the same function at all redshifts,
thus the degeneracy between the spectrophotometric correction and the
model fitting is partially broken because of the range of redshift
covered (0.03--0.25). Typically in our fitting we find this
spectrophotometric correction to be of the order 5--10\% change in the
response function of \citeauthor{lewis02}\ for the best-fitting models
(within 3$\sigma$).  These values represent the RMS relative
difference over the wavelength range between the polynomial correction
and a constant value (representing changes in the relative, not
absolute, spectrophotometry). Since the response function was measured
from standard stars observed during a single hour, such a difference
could arise from: (i) unaccounted for, reduction or throughput
discrepancies between extended and point sources, and/or; (ii)
variations in the instrument response over the \grs\ survey time.

\begin{figure*}[ht]
\centerline{
\epsfxsize=9.2cm
\epsfbox{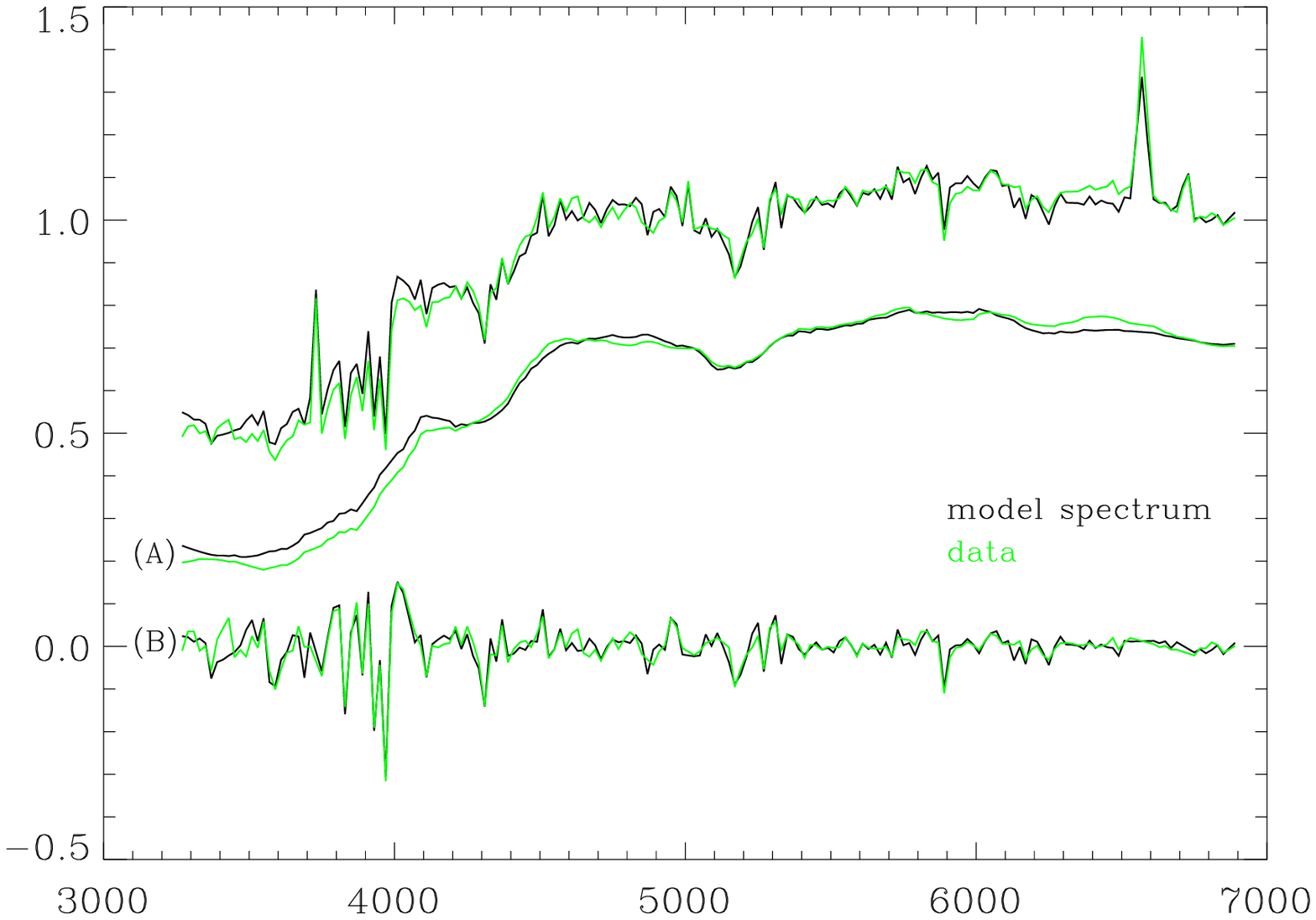}
\epsfxsize=9.2cm
\epsfbox{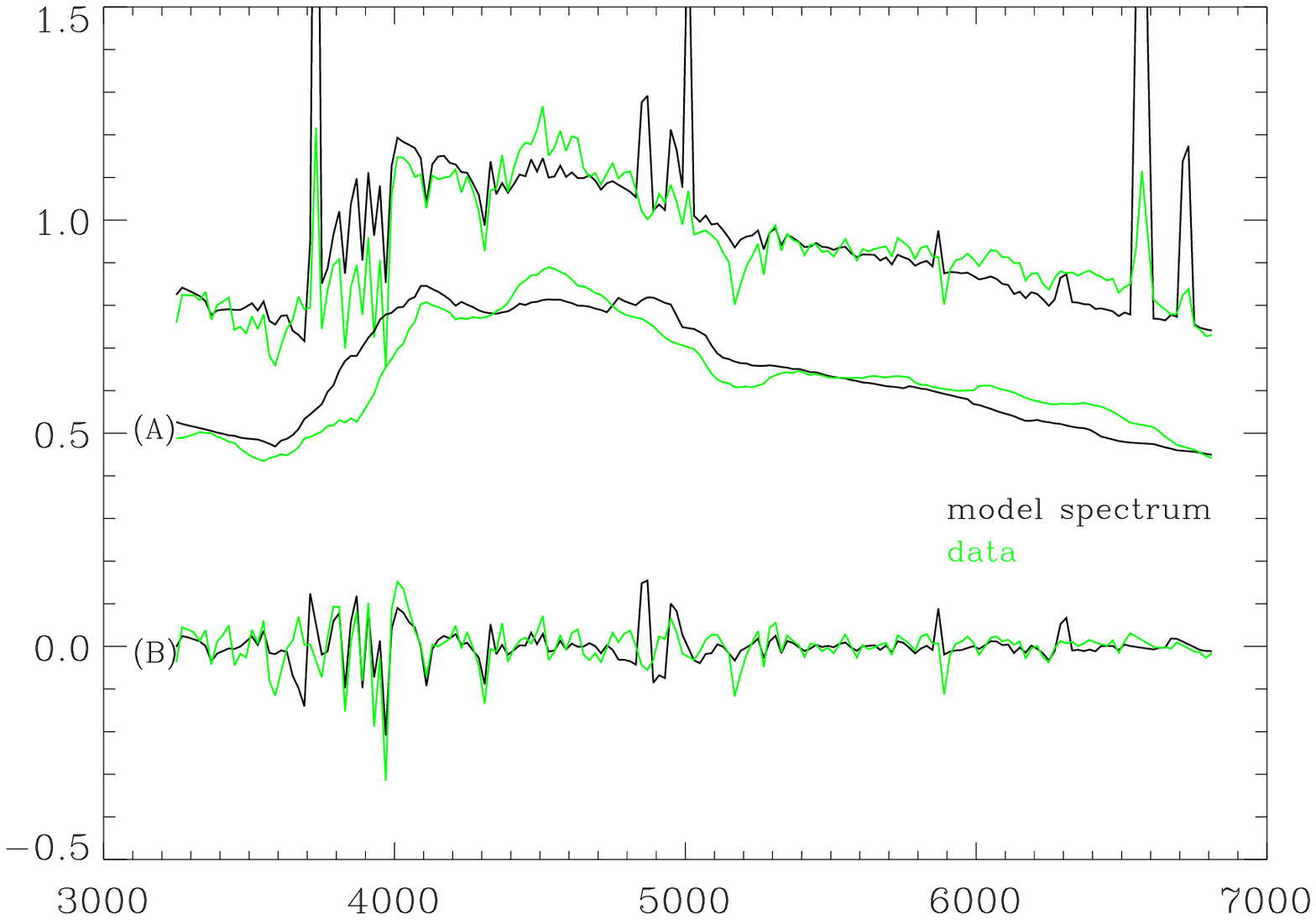}
}
\caption{Example model and data spectra{}
  with the low-pass (A) and high-pass (B) information separated.  The
  data have been adjusted for possible spectrophotometric errors as
  described in Section~\ref{sec:fit}.  FOM~A is the reduced
  \chisq\ value determined from the difference between the
  smoothed spectra (offset by $-$0.3 in the plot).  FOM~B is
  determined from the high-pass spectra (offset by $-$1.0). Note that
  the strongest emission lines were excluded from the fitting
  procedure.  The first plot shows a good fit with reduced \chisq\
  values of order unity for both figures of merit.  The second plot
  shows a poor fit for both FOMs ($\chi^2/\nu \sim 17$ and 6 for~A
  and~B, respectively).}
\label{fig:example-fitting}
\end{figure*}

In evaluating the fit between a model spectrum and a \grs\ data
spectrum, we compare the low-pass and high-pass information separately
to form two figures of merit (FOMs) from the normalized spectra.  This
is illustrated in Figure~\ref{fig:example-fitting}.  The normalization
is to set a mean of unity over the rest-wavelength range
4100--6200\AA, with the lower limit being just above the 4000\AA\ 
break and the upper limit being just in the detected range for
galaxies with $z=0.25$.  Both the normalized model spectrum and
normalized data spectrum are smoothed using a top-hat function of
width 200\AA\ (10 sampling points).  The original spectra are divided
by the smoothed spectra to produce the high-pass spectra. FOM~A is the
reduced \chisq\ value from the difference between the low-pass
spectra and FOM~B is is the reduced \chisq\ value from the
high-pass spectra.

This high-pass FOM~B is thus robust against any large-scale errors in
the spectrophotometry and in fact we find FOM~B is negligibly
different whether we include a spectrophotometric correction or not.
FOM~B places more reliable constraints on the best-fit star-formation
scenarios since it is less affected by the systematic uncertainties
of spectrophotometry and extinction. FOM~A is an independent check:
our aim is to get consistent histories from both the low-pass
(continuum) and high-pass (absorption line) information. From the 
best-fit star formation we can obtain a range of probable
spectrophotometric corrections that were used to determine the mean
and uncertainty in the \bj\ k-corrections of the galaxy spectra
\citep{madg02}.

In order to determine the uncertainties in our set of spectra we take
the empirical approach of dividing up the survey into ten separate,
approximately-equal sky areas. We compute the mean spectrum separately
for each area and use the variance between them to set out errors as a
function of wavelength.  The RMS errors computed this way are around
0.2--1\% for the high-pass spectrum and 1--3\% for the low-pass
spectrum.  An additional uncertainty of 1.5\% per wavelength bin is
added in quadrature to account for intrinsic model inaccuracies due
to, for example, the signal-to-noise ratio of the spectra in the input
atlas. This uncertainty value was chosen so that the reduced
\chisq\ values for the best-fit star-formation scenarios were
approximately unity.

For a given star-formation scenario, we average the FOM values over
all the redshift bins.  We calculate the figure-of-merit quantities by
summing over all wavelengths except near strong nebular emission lines
(O{\small II} 3727; O{\small III} 5007; H$\alpha$ 6563; N{\small II}
6583; S{\small II} 6716 \& 6730).  We do this because our goal is to
calculate the \sfh\ from the stellar emission only.  The nebular-line
emission is a measure of the instantaneous \sfr\ in H{\small II}
regions, but the PEGASE code uses over-simple assumptions to translate
\sfr\ to a set of line ratios.  Since understanding ionization levels
and interpreting line ratios are complex subjects, and since the
current strength of emission lines does not constrain past star
formation rates, we choose not to use this information.  In addition
the emission lines will be contaminated by active galactic nuclei in a
few percent of our galaxies, so ignoring the lines will minimize the
bias this causes.

\subsection{Results} \label{sec:results-sub}

\begin{figure*}[ht]
\centerline{
\epsfxsize=9.2cm
\epsfbox{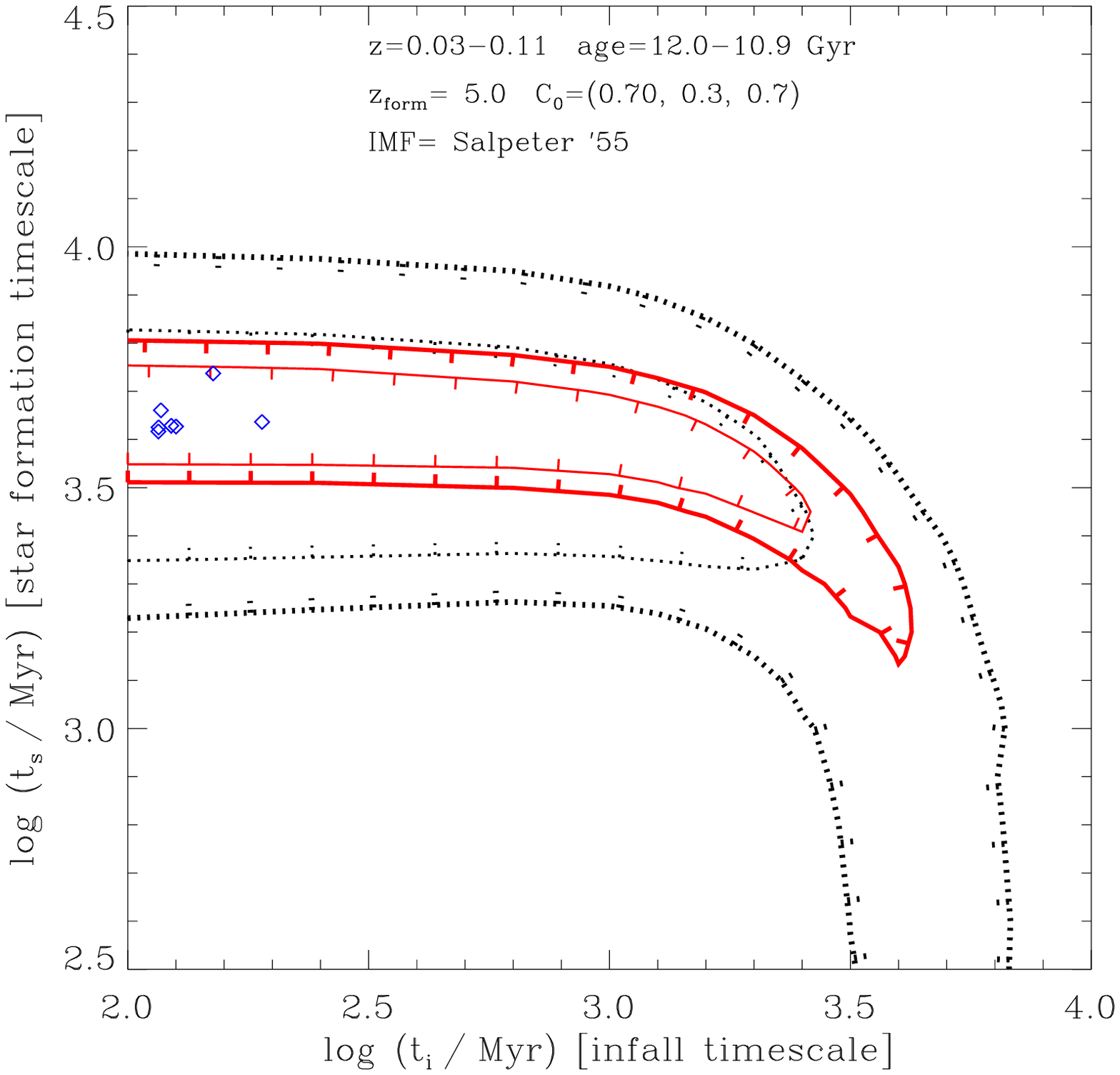}
\epsfxsize=9.2cm
\epsfbox{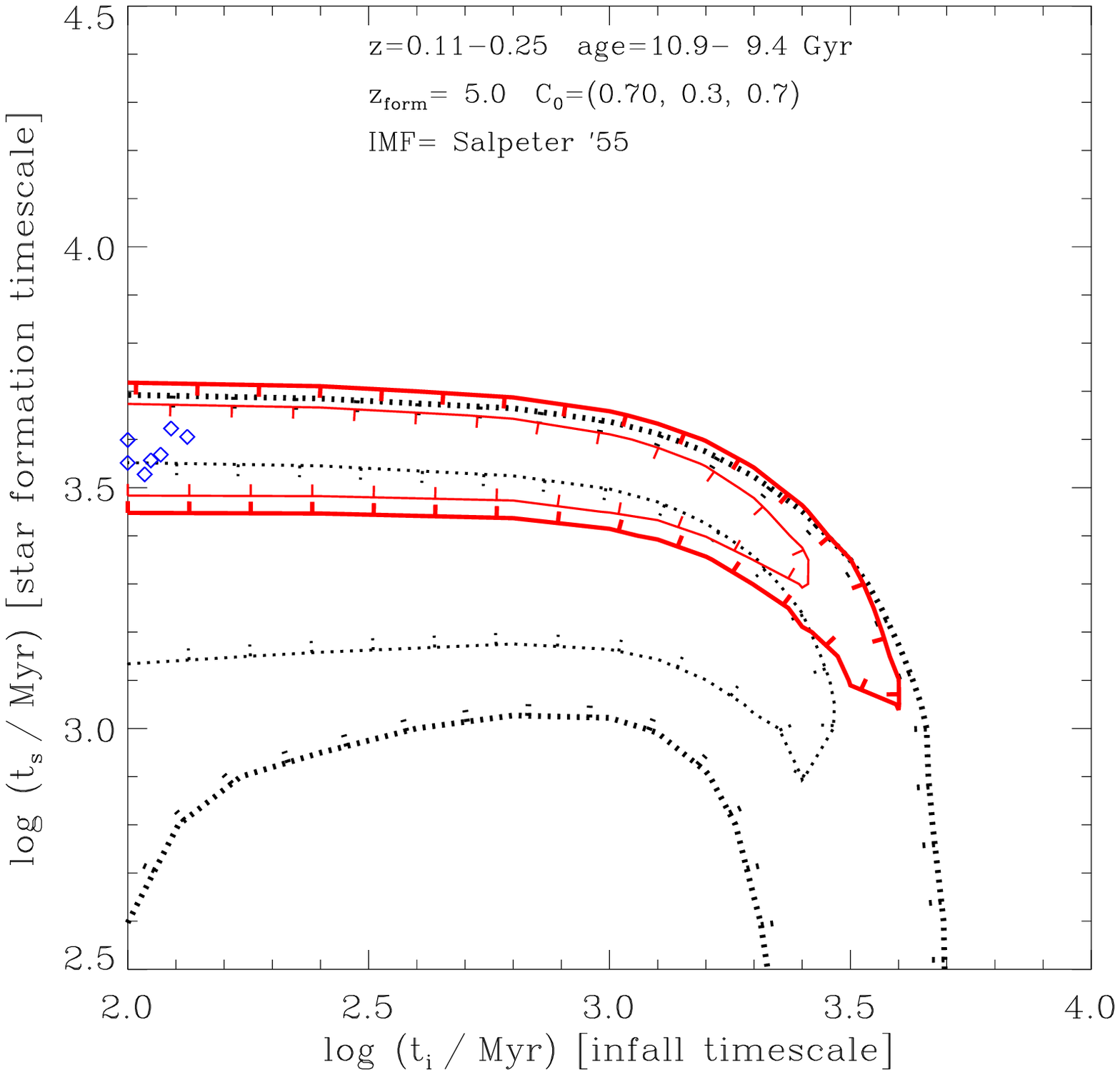}
}
\caption{Best-fit regions of $\log \tsf$ versus $\log \tin${}
  for cosmology $\cal C$1 with $\zform=5$.  The two plots use
  different redshift ranges from the data.  The contours represent the
  2$\sigma$ and 3$\sigma$ formal confidence limits, with dotted lines
  for FOM~A and solid lines for FOM~B. The FOMs were calculated using
  the average from 7 redshift bins at each grid point. The diamonds
  represent the FOM~B best-fit parameters for each bin.
  Section~\ref{sec:phys-param} describes the cosmic SFR
  parameterization using a star-formation timescale ($\tsf$) and an
  infall timescale ($\tin$).}
\label{fig:ti-versus-ts}
\end{figure*}

For the fitting procedure we calculate a grid of model values,
$(\tin,\tsf,\zform)$ for the physical parameterization and
$(\alpha,\beta,r)$ for the empirical parameterization.  We estimate
that after our data processing there are approximately 20 degrees of
freedom for FOM~A (about 22 independent wavelength points) and 200
degrees of freedom for FOM~B.  The minimum $\chi^2/\nu$ value was
about unity for both figures of merit using the uncertainties
described above.  We believe the additional uncertainty of 1.5\%
necessary to obtain a suitable \chisq\ value reflects the
remaining imprecision of the model spectra, as these types of models
were conceived originally to model galaxy colors. Improved,
higher-resolution models will be required in future work. We proceed
to interpret the star-formation laws of the best-fitting models. We
note that the best fit matches the data to about 2\% RMS, so in
absolute terms we are accounting for the volumetric light emission
very well.  The problem remains that the high-pass data are much
better than the current state of the art in population synthesis
modeling.

In order to estimate the confidence limits on the FOMs ($\Delta \chi^2
/ \nu$), we investigated the variation of the best-fit parameters
using Monte Carlo simulations. It is not sufficient to apply the
standard confidence limits because of correlated uncertainties between
different wavelength points.  The investigation included using the
spectra from the ten positional bins separately, applying random
errors of 1.5\%, using different redshift ranges (e.g., 0.03--0.11,
0.11--0.25) and varying the dust extinction in the models.  As
expected, FOM~B is fairly robust and we use $\Delta \chi^2 / \nu =
0.15$ for the 3$\sigma$ (99.73\%) confidence limit. FOM~A is strongly
affected by using different positions on the sky and different
redshift ranges.  This is principally due to spectrophotometric
uncertainties, and we use $\Delta \chi^2 / \nu = 2.1$ for the
3$\sigma$ limit (when plotting contours using the full redshift
range). In neither case is the FOM significantly affected by changes
in the dust extinction used in the model. This is not surprising for
FOM~B since it is determined from high-pass spectra.  A
spectrophotometric adjustment is applied to the data before
determining the FOMs and therefore FOM~A is principally sensitive to
variations over ranges of 200\AA\ (the smoothing length), such as the
4000\AA\ break, and is less sensitive to extinction.

\begin{figure*}[ht]
\centerline{
\epsfxsize=9.2cm
\epsfbox{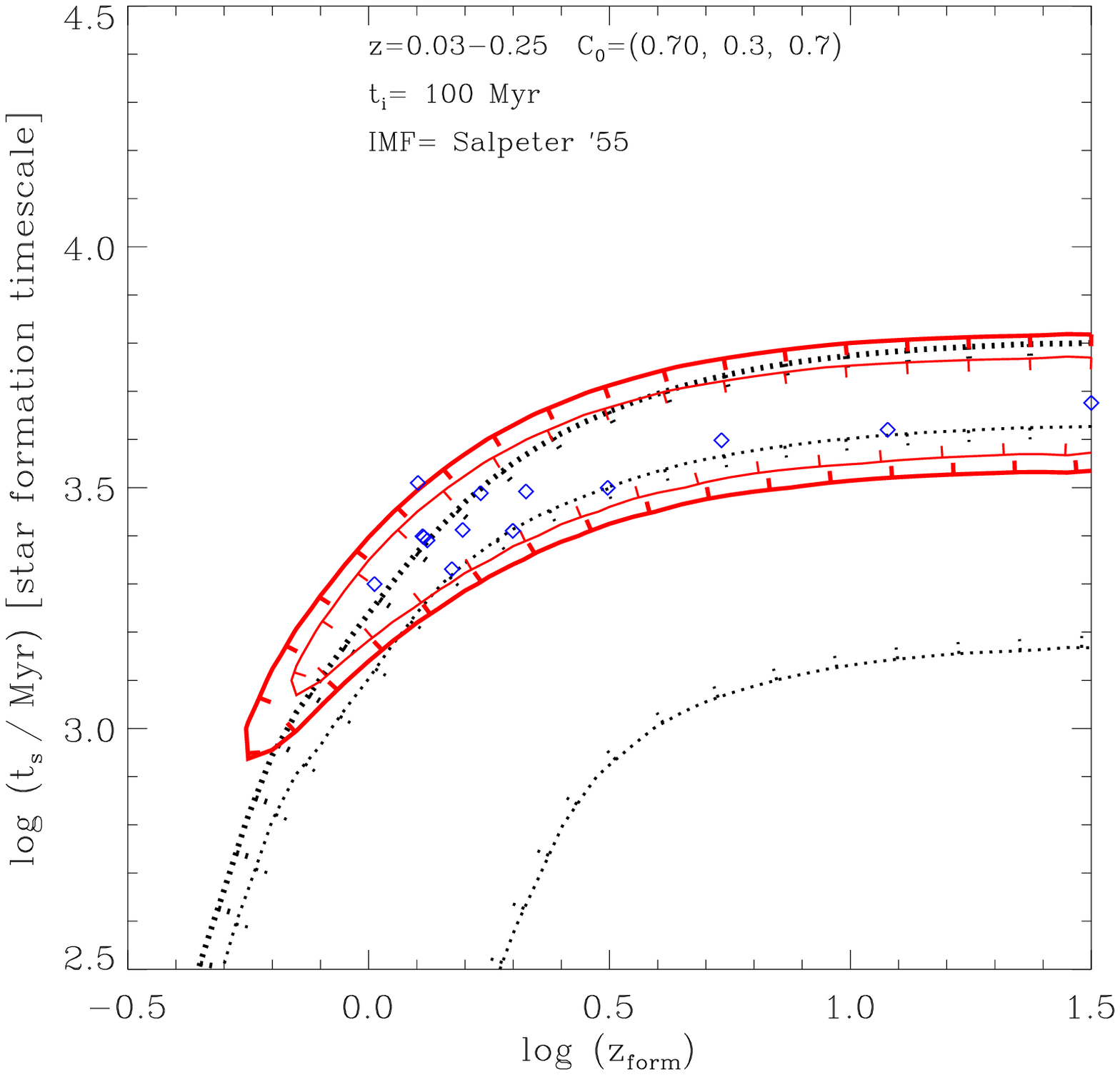}
\epsfxsize=9.2cm
\epsfbox{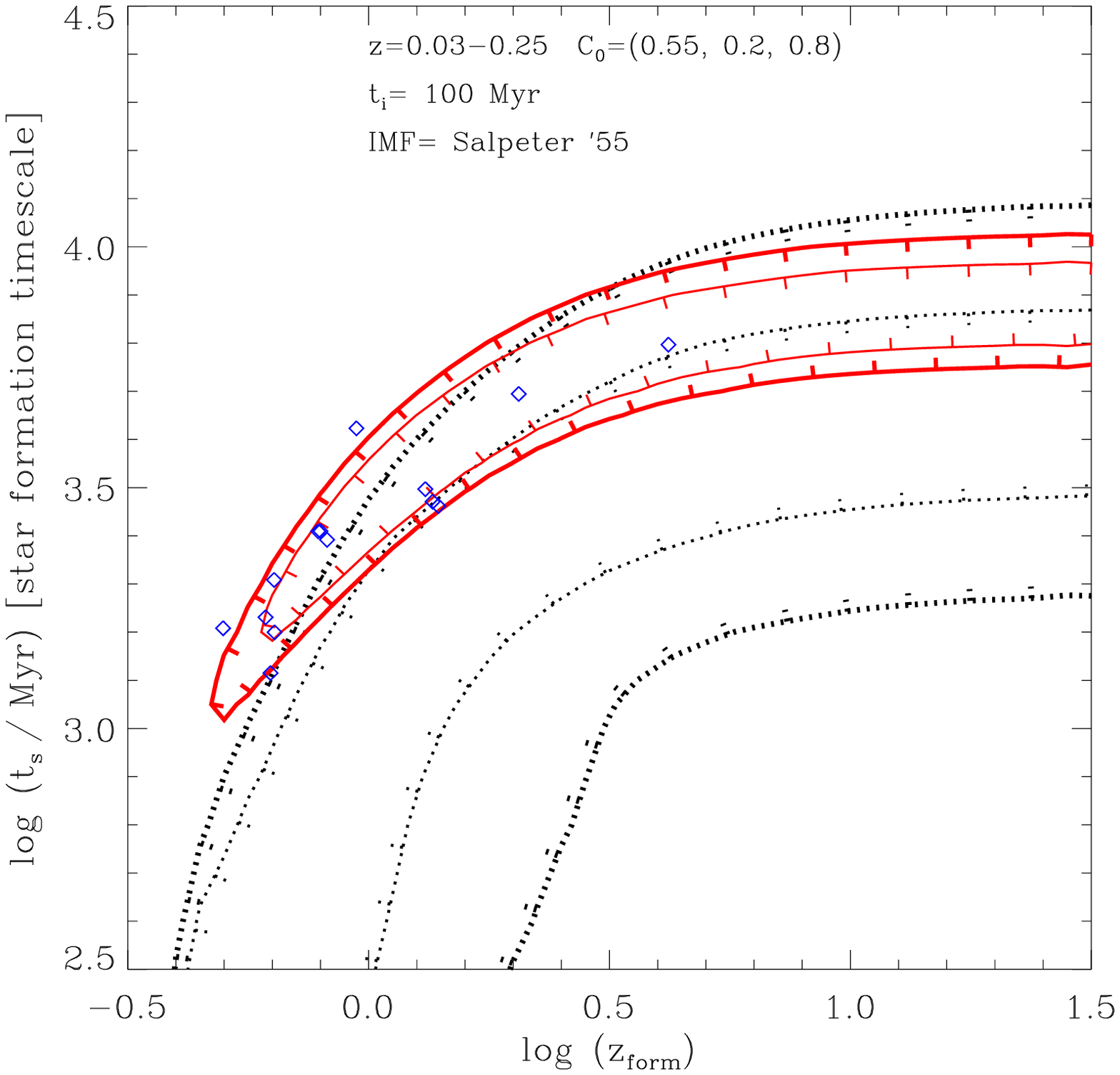}
}
\caption{Best-fit regions of $\log\tsf$ versus $\log\zform${}
  for two different cosmologies. See Figure~\ref{fig:ti-versus-ts} for
  contour meanings.}
\label{fig:zform-versus-ts}
\end{figure*}

\begin{figure*}[ht]
\centerline{
\epsfxsize=9.2cm
\epsfbox{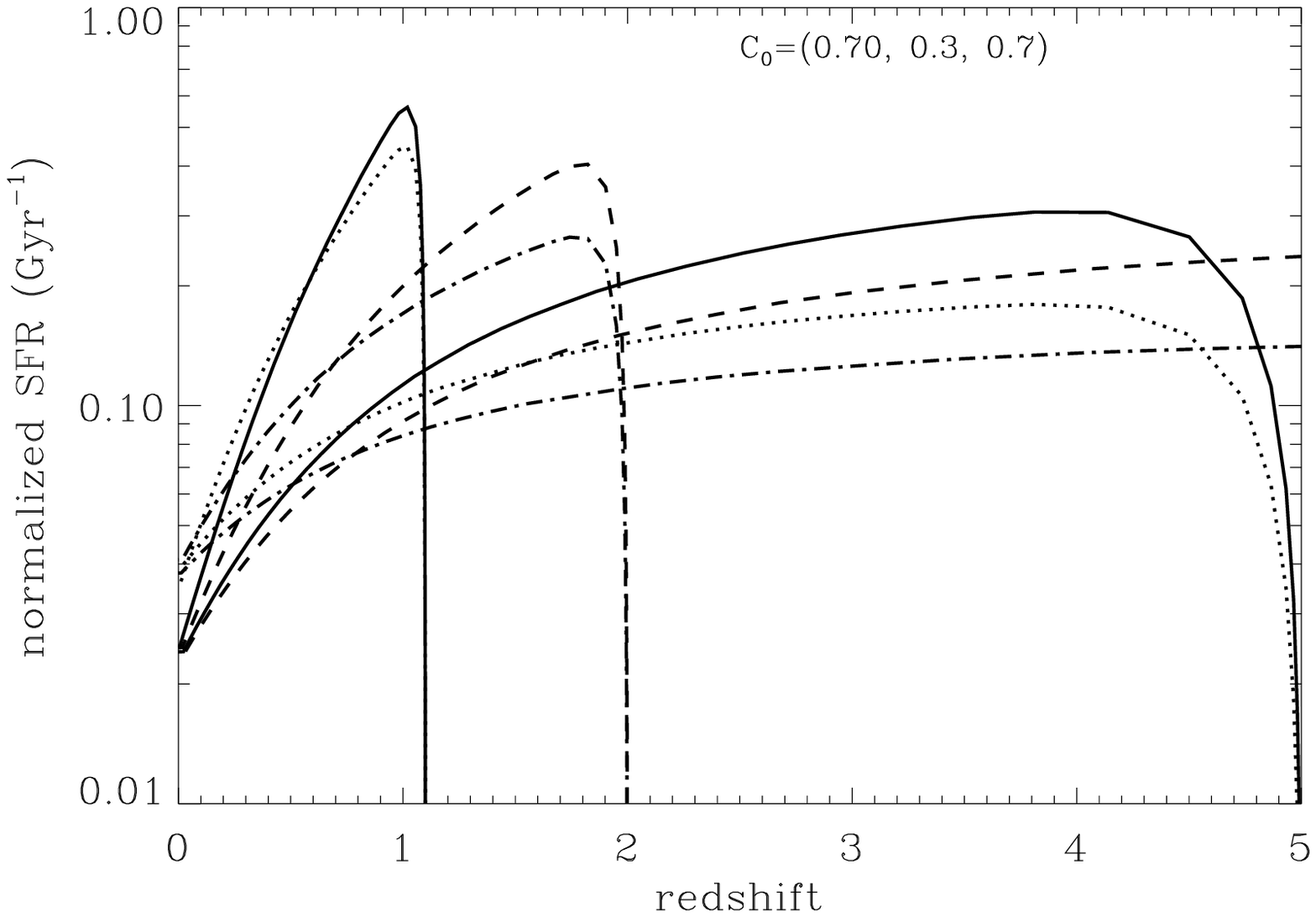}
\epsfxsize=9.2cm
\epsfbox{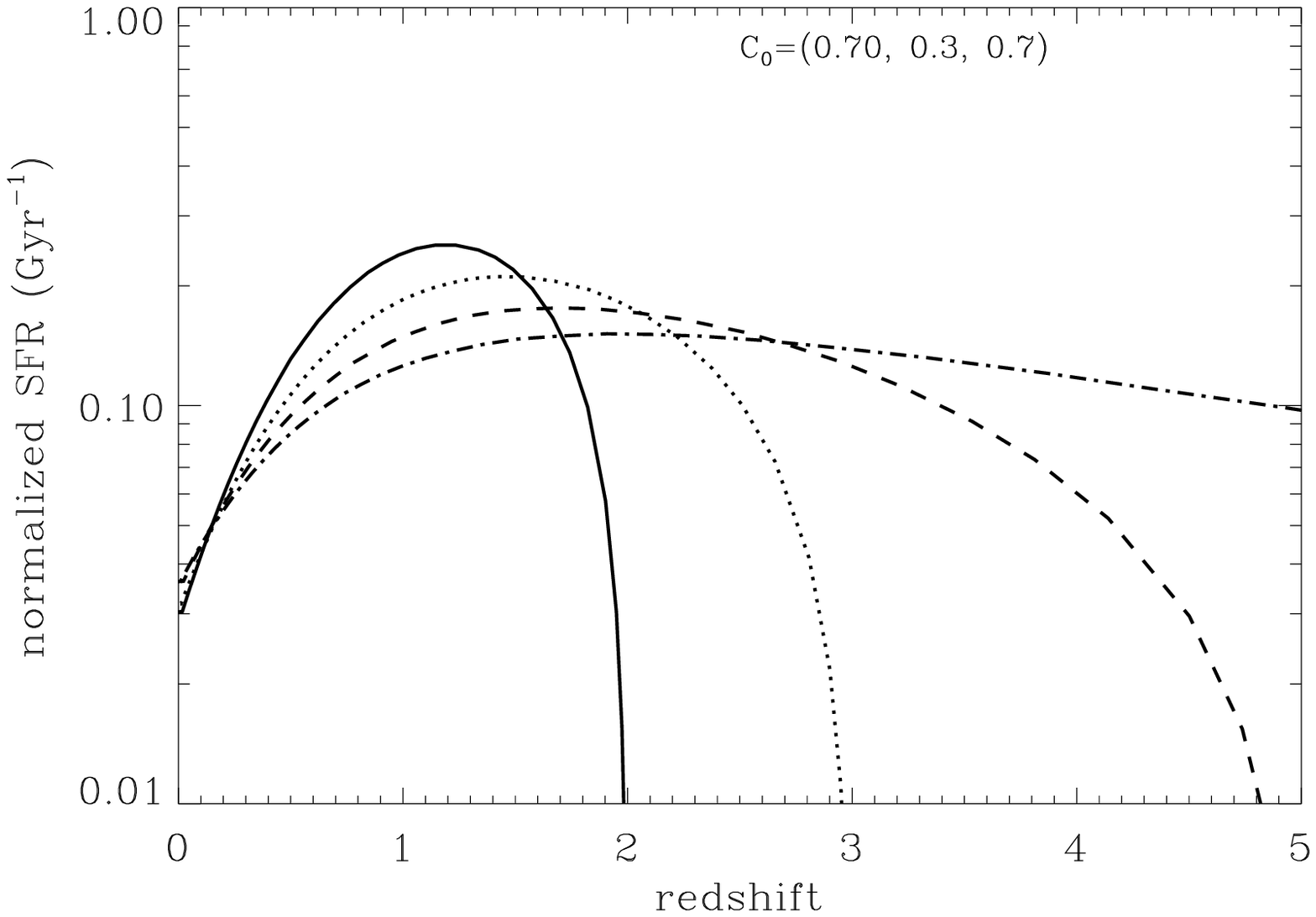}
}
\caption{Example set of star-formation scenarios{}
  that are within 3$\sigma$ of the minimum FOM values for cosmology
  $\cal C$1.  The first plot shows scenarios with $\tin \ll \tsf$ and
  the second plot with $\tin \sim \tsf$.}
\label{fig:best-fit-scenarios}
\end{figure*}

The best-fit parameters are fairly degenerate.
Figure~\ref{fig:ti-versus-ts} shows the best-fit regions in $\log
\tsf$ versus $\log \tin$ space with $\zform=5$.  The results for the
lower and higher redshift ranges are shown separately.  The solid
contours represent the 95.4\% and 99.73\% (`2$\sigma$' and
`3$\sigma$') confidence boundaries for FOM~B, while the dotted
contours are for FOM~A.  Notably, the regions contained by the FOM~B
boundaries at low and high redshift are similar whereas the FOM~A
boundaries are significantly different. We do not over-interpret this,
noting that FOM~A is less reliable due to spectrophotometric
uncertainties. The best-fit model (FOM~B) has $\tsf \sim 4000$\,Myr
with $\tin \la 200$\,Myr (note degeneracy for low $\tin$). Both
redshift ranges have $\tin \la \tsf$ within the 2$\sigma$ levels even
though the SFR is similar on interchange of $f\tin$ and $\tsf$ (see
approximation, Eqn.~\ref{eqn:sfr-model}). However, the metallicity
evolution is different depending on whether infall of new gas or the
star-formation timescale determines the SFR at late times.

The figure-of-merit values become degenerate for $\tin \ll \tsf$
(galaxies form quickly in comparison with the star-formation
timescale) while maintaining a good fit. In
Figure~\ref{fig:zform-versus-ts}, $\tin$ is set at 100\,Myr and the
best-fit regions of $\log\tsf$ versus $\log\zform$ are identified for
cosmologies $\cal C$1 and $\cal C$2. At the 3$\sigma$ limits (FOM~A
and~B), the redshift of formation is greater than or about 0.65.

To aid understanding of the meaning of our results,
Figure~\ref{fig:best-fit-scenarios} shows a variety of star-formation
histories that are within the 3$\sigma$ limits. The first plot shows
scenarios with $\tin \ll \tsf$, i.e., identified from
Figure~\ref{fig:zform-versus-ts} ($\cal C$1).  The second plot
(Fig.~\ref{fig:best-fit-scenarios}) shows `smoother' scenarios with
$\tin \sim \tsf$, identified from plots similar to those shown in
Figure~\ref{fig:ti-versus-ts}, the right edges of the 3$\sigma$
boundaries.

\begin{figure*}[ht]
\centerline{
\epsfxsize=9.2cm
\epsfbox{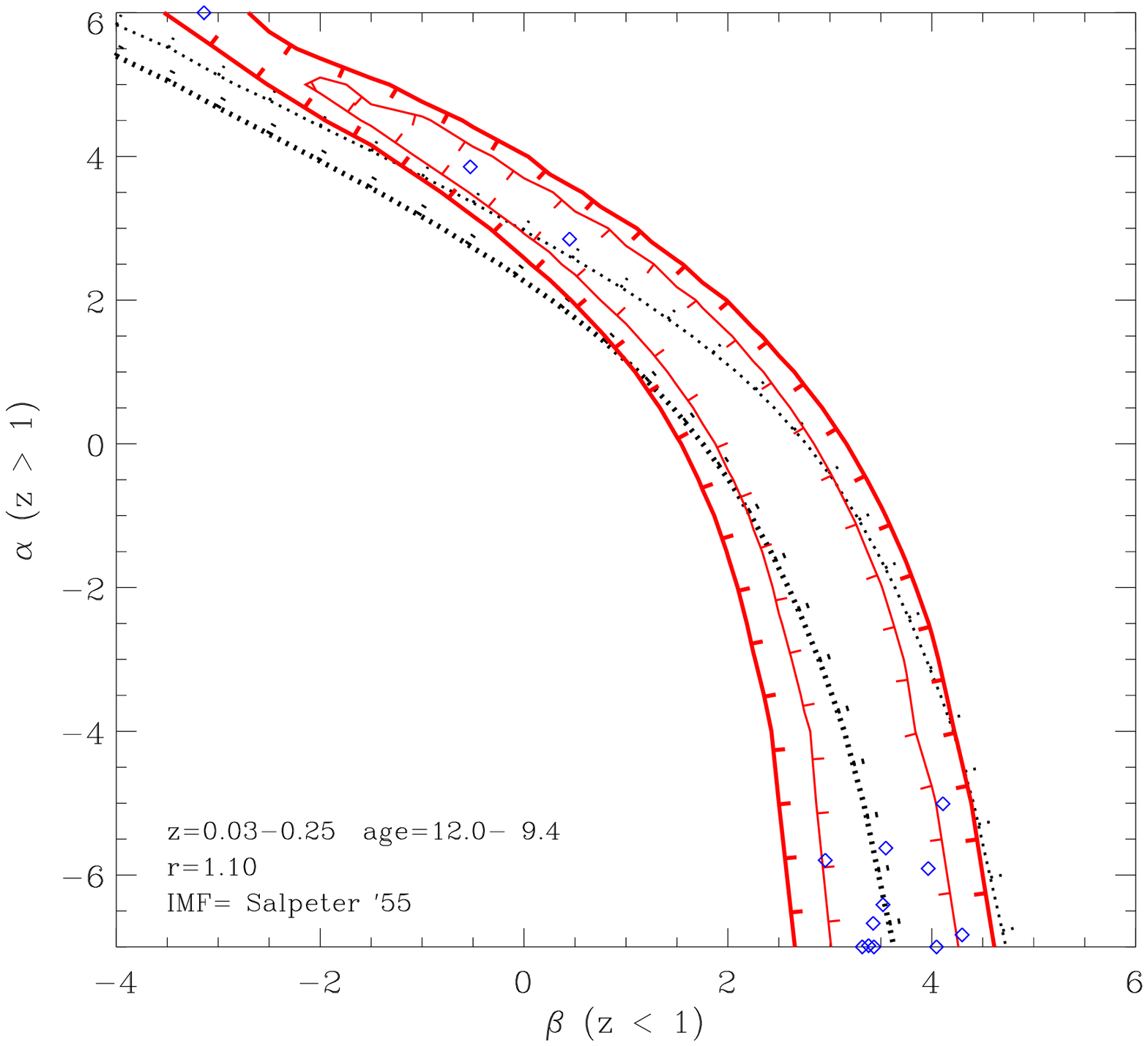}
\epsfxsize=9.2cm
\epsfbox{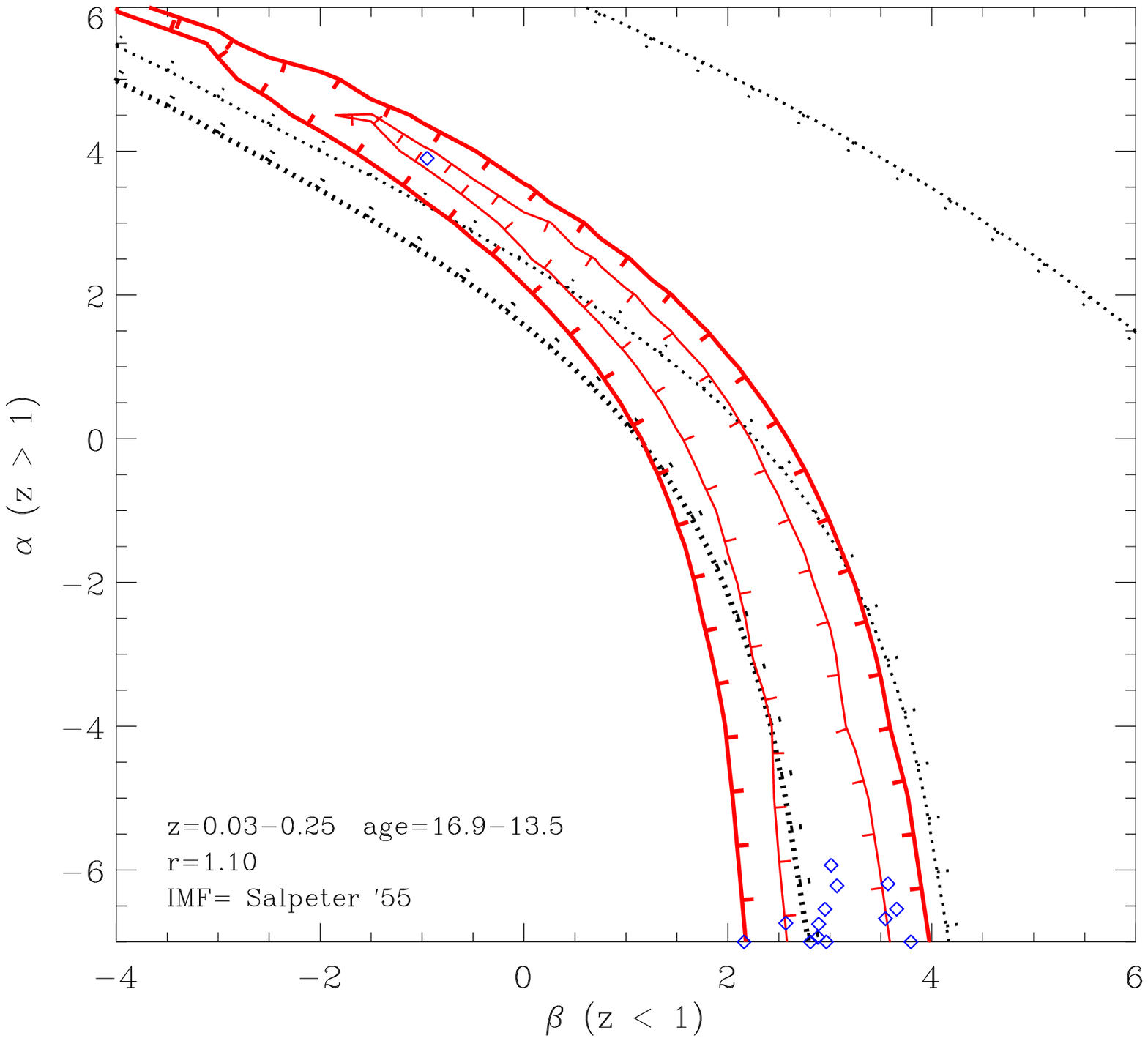}
}
\caption{Best-fit regions of $\alpha$ versus $\beta${} 
  for cosmologies $\cal C$1 and $\cal C$2. See
  Figure~\ref{fig:ti-versus-ts} for contour meanings.
  Section~\ref{sec:emp-param} describes the cosmic SFR
  parameterization with $\sfrate\propto(1+z)^\beta$ for $0<z<1$ and
  $\sfrate\propto(1+z)^\alpha$ for $1<z<5$.}
\label{fig:alpha-versus-beta}
\end{figure*}

\begin{figure*}[ht]
\centerline{
\epsfxsize=9.2cm
\epsfbox{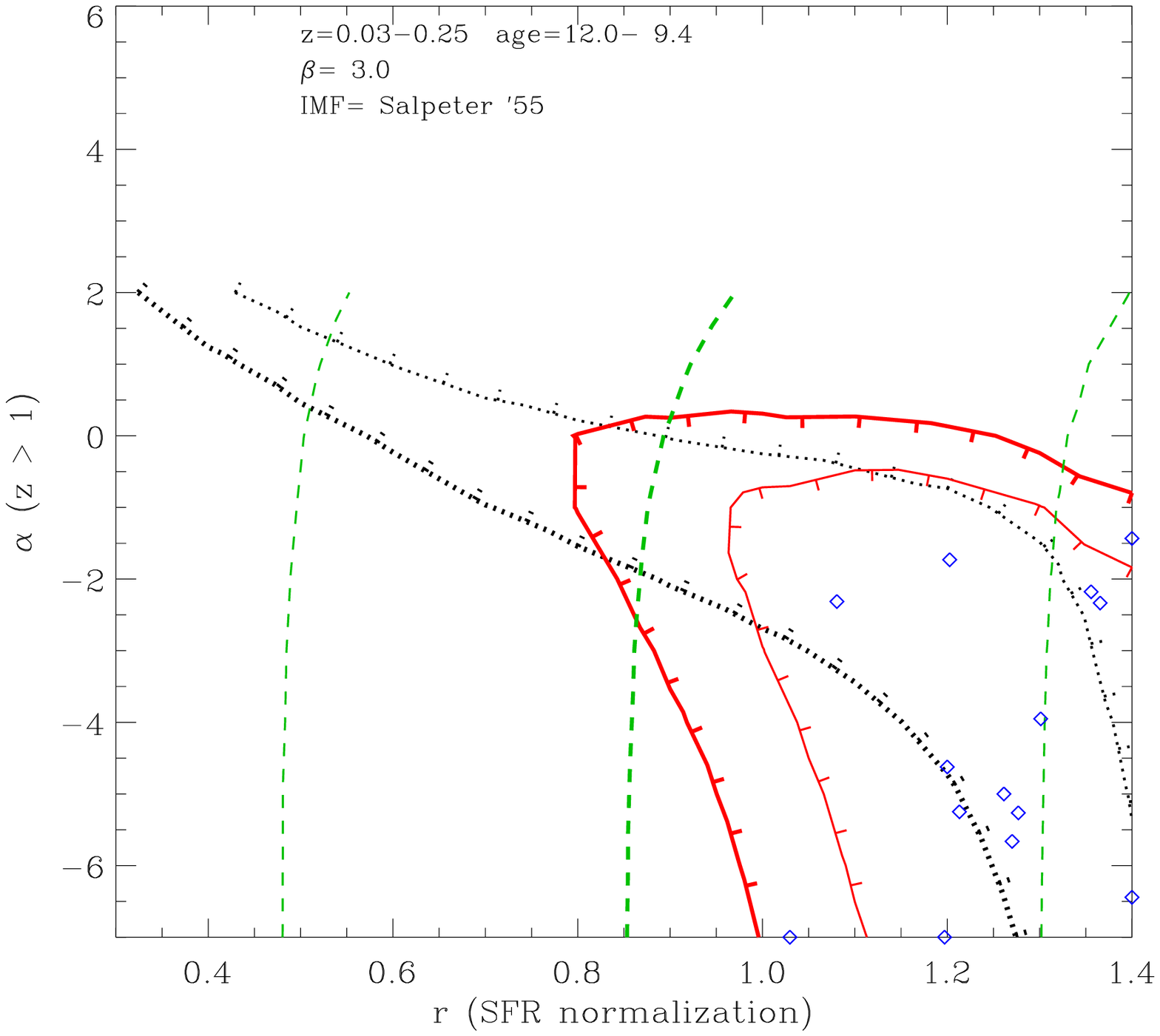}
\epsfxsize=9.2cm
\epsfbox{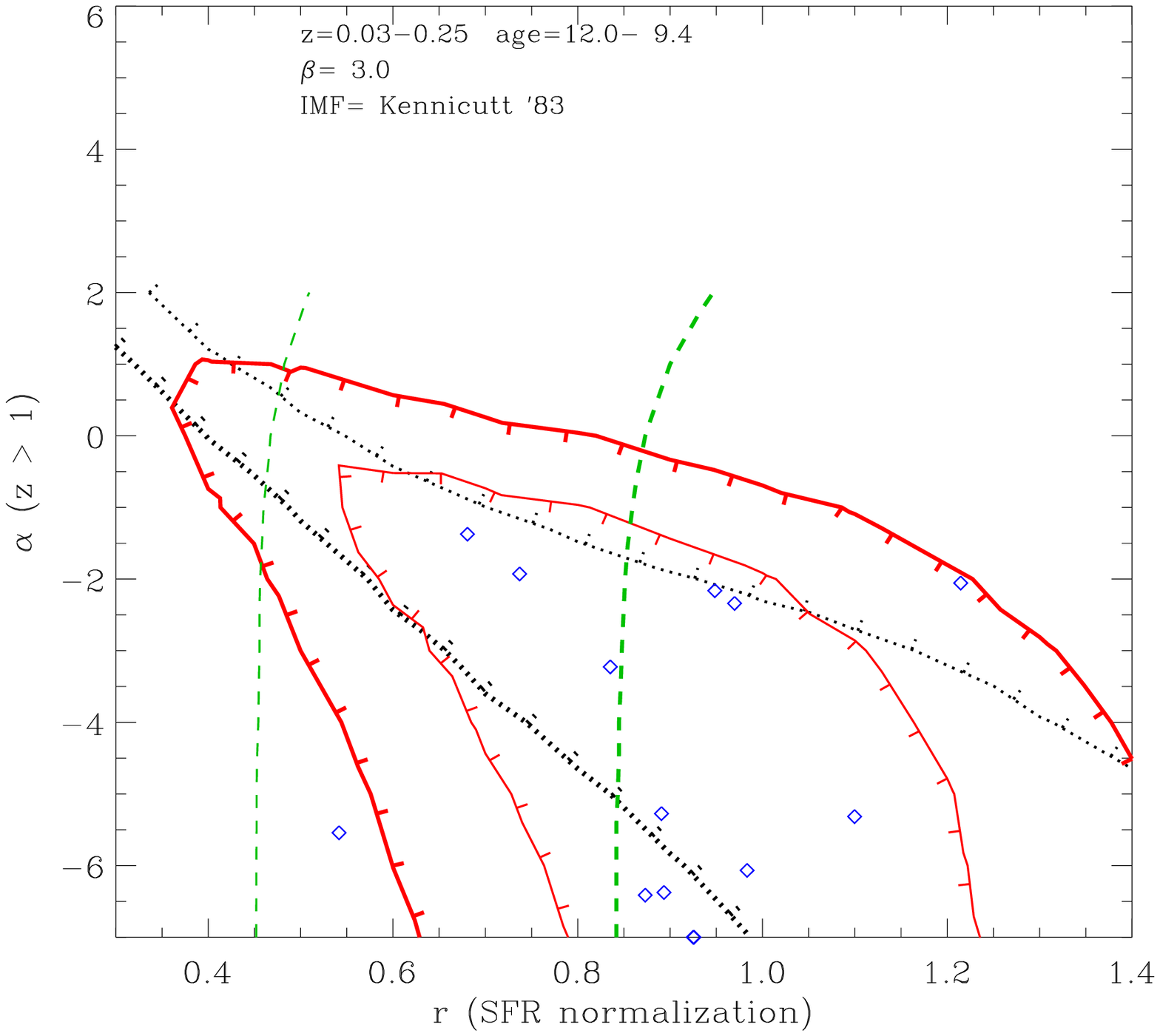}
}
\caption{Best-fit regions of $\alpha$ versus $r$ with $\beta=3${}.
  The two plots are for different IMFs with cosmology $\cal C$1.  See
  Figure~\ref{fig:ti-versus-ts} for contour meanings, in addition, the
  vertical dashed lines show the final metallicity averaged on the
  luminosity: from left to right, $Z=0.01$, 0.02 and 0.04 in the first
  plot, and; 0.01 and 0.02 in the second plot.}
\label{fig:alpha-versus-r}
\end{figure*}

\begin{figure*}[ht]
\centerline{
\epsfxsize=9.2cm
\epsfbox{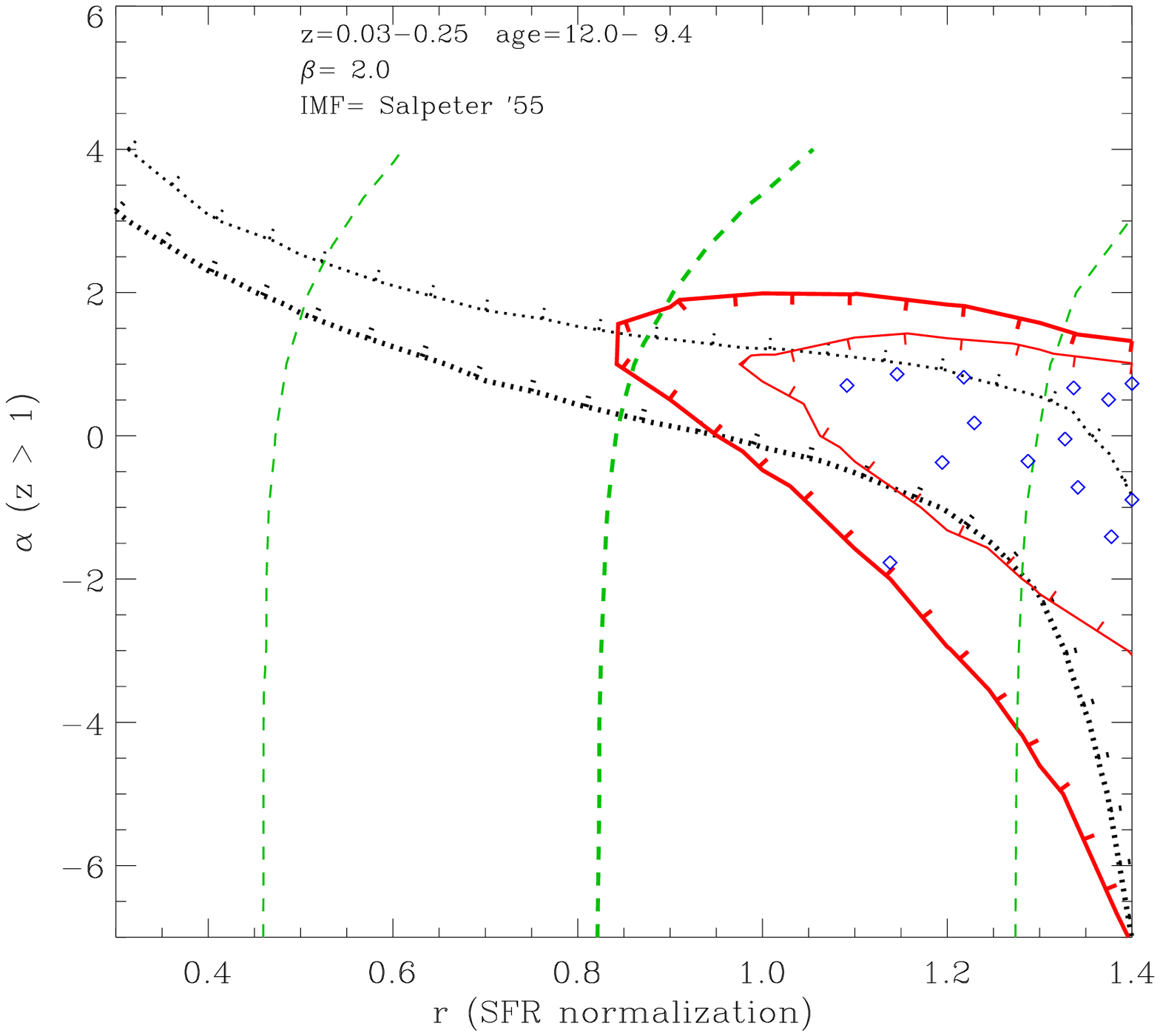}
\epsfxsize=9.2cm
\epsfbox{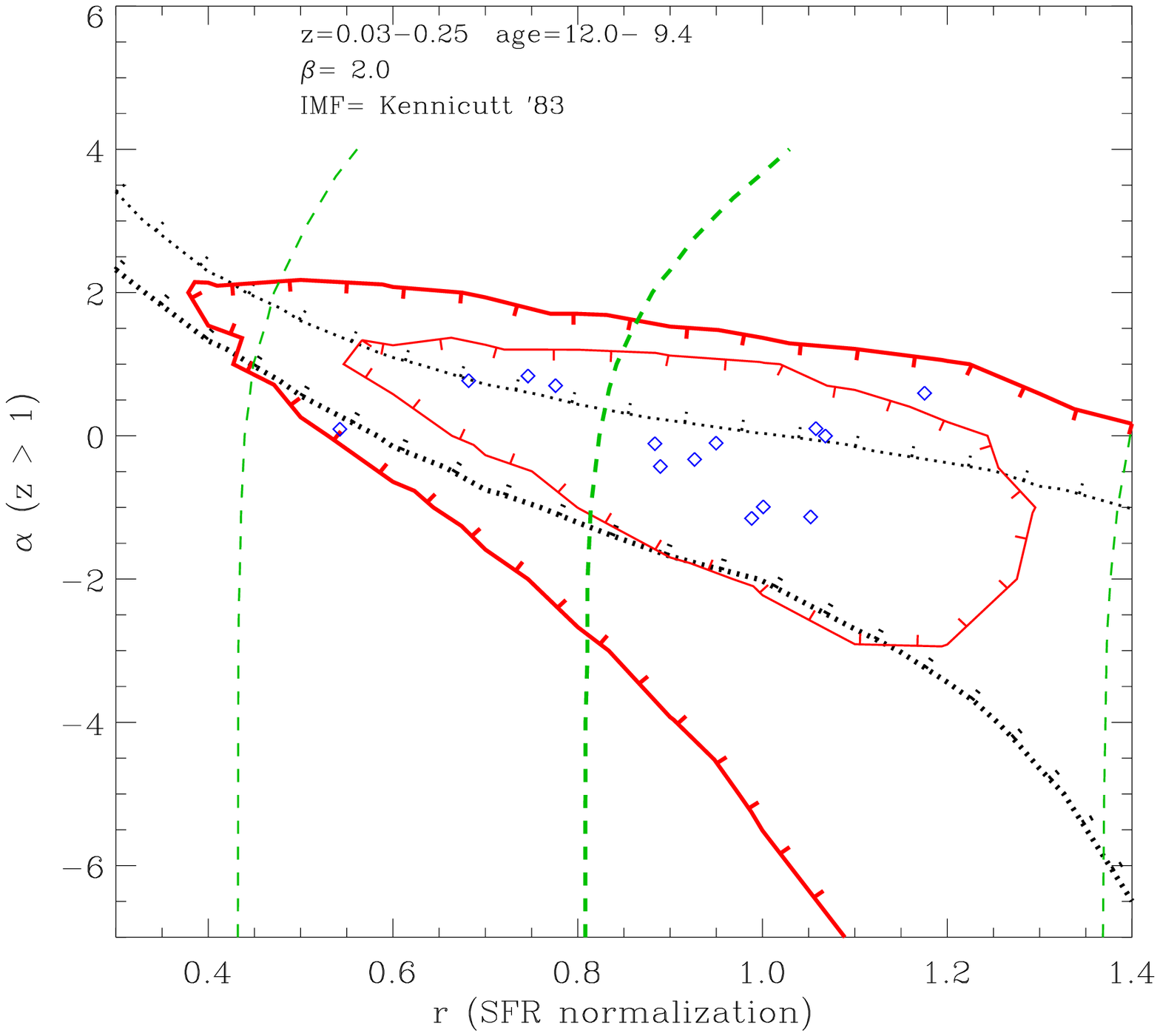}
}
\caption{Best-fit regions of $\alpha$ versus $r$ with $\beta=2$.}
\label{fig:alpha-versus-r-2}
\end{figure*}

The best-fit scenarios cover a large range of $\zform$ but have the
following in common: the normalized SFR at $z=0$ is in the range
0.02--0.04\,Gyr$^{-1}$ and 84--92\% of stars formed prior to
$z=0.3$ for $\cal C$1.

It can be seen that the main constraint is on the slope of the \sfr\ 
for $0<z<1$.  Fitting a SFR of $(1+z)^{\beta}$ from $z=0$ to 1 to the
{\em natural} scenarios with $\zform > 1$, we obtain from the
3$\sigma$ confidence limits: $1 \la \beta \la 4.5$ ($\cal C$1) and $1
\la \beta \la 4$ ($\cal C$2).  If we fit a SFR of $(1+z)^{\alpha}$
from $z=1$ to 4, there is no lower limit on $\alpha$ since at least
one of the well-fitting models has little or no star formation prior
to $z=1$.  The upper limit from these natural scenarios is $\alpha \la
1$ for both cosmologies.

The $\tin$-$\tsf$-$\zform$ parameterization is limited in its scope
for changes of SFR with time. Therefore, to further test \sfh, we look
at the $\alpha$-$\beta$-$r$ parameterization.
Figure~\ref{fig:alpha-versus-beta} shows best-fit regions in $\alpha$
versus $\beta$ with $r=1.1$ for both cosmologies.

In these scenarios, the galaxies start fully constituted with gas
(there is no infall) and consistent evolution of the metallicity is
implemented. Constant metallicity scenarios were also tested but found
not to be consistent within the 3$\sigma$ limit on FOM~B.  If we take
a SFR normalization of 1.1, meaning a total mass of stars formed
between $z=5$ and $z=0$ equal to 1.1 times the mass of gas available,
there is a degeneracy across the plane of $\alpha$ versus $\beta$
(Figure~\ref{fig:alpha-versus-beta}).  Scenarios with $\beta<0$ cannot
be ruled out for $\alpha>2.5$.  However, this would imply a {\em
  minimum} in the SFR around $z=1$ which is in disagreement with many
cosmic SFR density studies based on photometry.  These also represent
a branch of solutions which are impossible to produce using the
physical parameterization.  The conservative upper limit of this study
is $\beta<5$.  Taking the conservative lower limit on $\beta$ from the
analysis by \citet{Hogg02}, $\beta>1.3$, we obtain $\alpha<3$ (for
$\cal C$1). If we assume $\alpha\la0$ (i.e., star formation declined
or remained constant for $z>1$ which corresponds to the lower-right
branch of solutions in Figure~\ref{fig:alpha-versus-beta}) then we
obtain a range of $1.5<\beta<5$. This is in very good agreement with
\citeauthor{Hogg02}'s results which represent an ensemble of different
luminosity-density based indicators.

Figure~\ref{fig:alpha-versus-r} shows best-fit regions in $\alpha$
versus $r$ with $\beta=3$ for $\cal C$1.  The two plots show the
results for different IMFs. There is no lower limit on $\alpha$ given
$\beta=3$. There are upper limits of $\alpha<0.5$ for the
\citeauthor{Salp55} IMF and $\alpha\la1$ for \citeauthor{Kenn83} IMF,
with $\alpha<0$ at the 2$\sigma$-confidence level for both IMFs.
There is some degeneracy between the final metallicity ($z=0$) and the
IMF.  For example, if we take star-formation scenarios within
$\alpha>-1$ and the 3$\sigma$ confidence boundaries, there are
scenarios with around solar to twice-solar metallicity using the
\citeauthor{Salp55} IMF and with around half-solar to solar
metallicity using the \citeauthor{Kenn83} IMF.

With $\beta\ga3$, the best-fit models have a decrease in SFR prior to
$z=1$ (2$\sigma$ limit of FOM~B).  Figure~\ref{fig:alpha-versus-r-2}
shows best-fit regions in $\alpha$ versus $r$ with $\beta=2$ for $\cal
C$1. Here, the best-fit models are consistent with a plateau in SFR
prior to $z=1$ or a marginal increase or decrease. An upper limit
would be $\alpha<1.5$ (2$\sigma$), if $\beta>2$ as is suggested by
most direct methods of tracing the cosmic SFR \citep{Hogg02}.

\section{Possible selection effects and biases} \label{sec:biases}

\begin{figure*}[ht]
\centerline{
\epsfxsize=9.2cm
\epsfbox{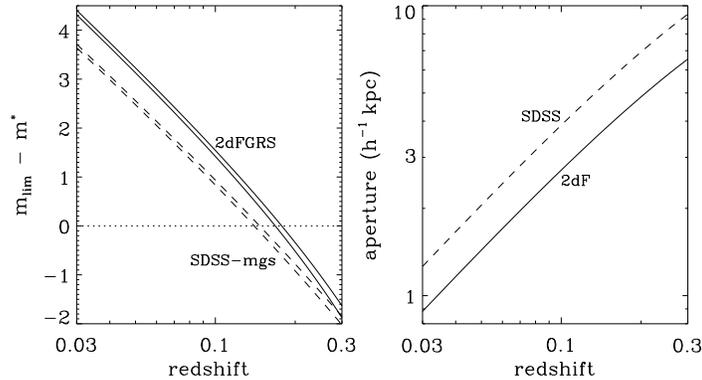}
}
\caption{Selection and aperture effects in the \grs\ and, for
  comparison, the SDSS main galaxy sample.
  The left plot shows the limiting magnitude relative to the
  characteristic magnitude of the \citet{Schech76} luminosity function
  versus redshift.
  For the \grs\ (\bj): $m_{\rm lim} \approx 19.45$;
  $m^{*} \approx -19.79 + 5 \log D_{\rm L} + 1.9z + 2.7z^2$ \citep{madg02};
  the solid lines show the limits with $\pm 0.04$ error in $M^{*}$ and
  $\pm 10$\% error in the k-correction.
  For the SDSS ($r'$): $m_{\rm lim} \approx 17.7$;
  $m^{*} \approx -20.83 + 5 \log D_{\rm L} + 2.5 \log (1+z)$ \citep{blanton01};
  the dashed lines show the limits with $\pm 0.03$ error in $M^{*}$
  and $\pm 20$\% error in the k-correction. The luminosity distance is
  calculated assuming a world model with $C_0 =$\,(1.0,0.3,0.7).
  The right plot shows the aperture scale
  for fiber diameters of $2''\!.1$ and $3''\!.0$.}
\label{fig:sel-ap}
\end{figure*}

All spectroscopic information is subject to the effect of aperture
bias \citep*[e.g., see the discussion of][]{KPF02}.  In particular, the
aperture may be too small to encompass a representative fraction of a
galaxies light. This effect will obviously be greater at lower
redshift.

Figure~\ref{fig:sel-ap} shows the increase in the effective size of a
2dF $2''\!.1$-fiber aperture with redshift. At $z=0.1$ this is
2.7\,$h^{-1}$\,kpc which is comparable to a typical large disk scale
length of 3\,$h^{-1}$\,kpc \citep{JL00}. In practice, the
effective aperture is more like 3.4\,$h^{-1}$\,kpc ($2''\!.6$) as the
median observation seeing was about $1''\!.5$. Thus it is reasonable
that the effect of aperture bias will be much smaller for $z>0.1$ as
we are sampling more than half the total light of galaxies.

\begin{figure*}[ht]
\epsfxsize=9.2cm
\centerline{
\epsfbox{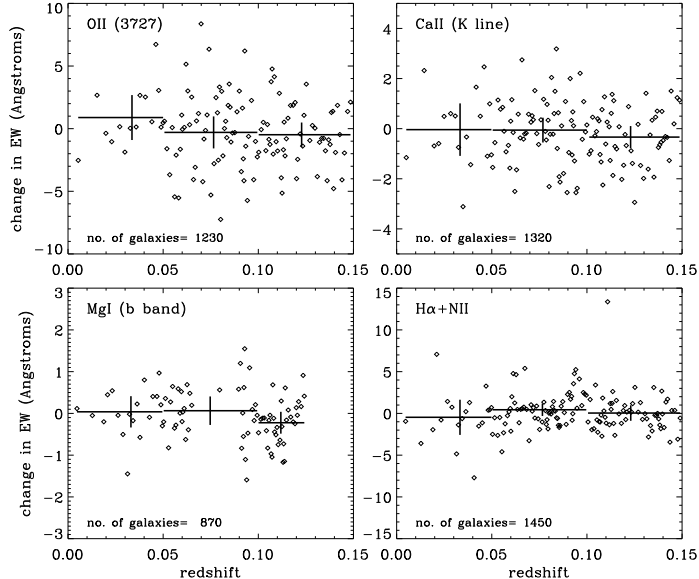}
}
\caption{Change of line equivalent widths{}
  -- O{\tiny II}, Ca{\tiny II}, Mg{\tiny I}, H$\alpha$ -- between
  measurements taken in poor seeing and in good seeing.  Each diamond
  represents the median change of a subsample of ten galaxies. The
  lines represent the mean of the subsamples, $\pm3\times$ the
  standard-error of the mean, divided into three redshift ranges.}
\label{fig:change-ew-lines}
\end{figure*}

The data allow for a further test of this because of its own internal
seeing variations. Spectra taken in bad atmospheric seeing are a rough
proxy for spectra taken through a larger aperture, as the object is
smeared out over a disk about the size of the seeing disk.  As a test
of this, we chose a sample of about 1500 galaxies ($z<0.15$) that had
measured spectra taken both in relatively good seeing ($\la1''\!.5$)
and in poor seeing ($\ga3''$) such that the difference in
representative aperture was greater than a linear factor of 1.5 for
each galaxy with $Q\ge3$ for both spectra. With this sample, we
measured the change in equivalent width (EW) of a number of lines
between the larger representative aperture and the smaller aperture.
In some individual spectra, the sky subtraction was inadequate for
accurate EW measurement.  Therefore, measurements for which the
reduced counts were too low or contaminated by sky-emission lines were
excluded. For further robustness, the galaxies were divided into
subsamples of ten as a function of redshift and the median change in
EW of each group was determined.  The results are shown in
Figure~\ref{fig:change-ew-lines}.  Similar results were obtained if a
weighted mean (by counts) was used rather than a median.

There is {\em no evidence} for any aperture effect in terms of the
average change of EW for the galaxies in this sample given a
difference in representative aperture with a linear factor in the
range 1.5--2.3.  We cannot rule out a small aperture effect on the
measured cosmic spectra particularly for $z<0.05$ where the test
sample is small.

\begin{figure*}[ht]
\epsfxsize=15.0cm
\centerline{
\epsfbox{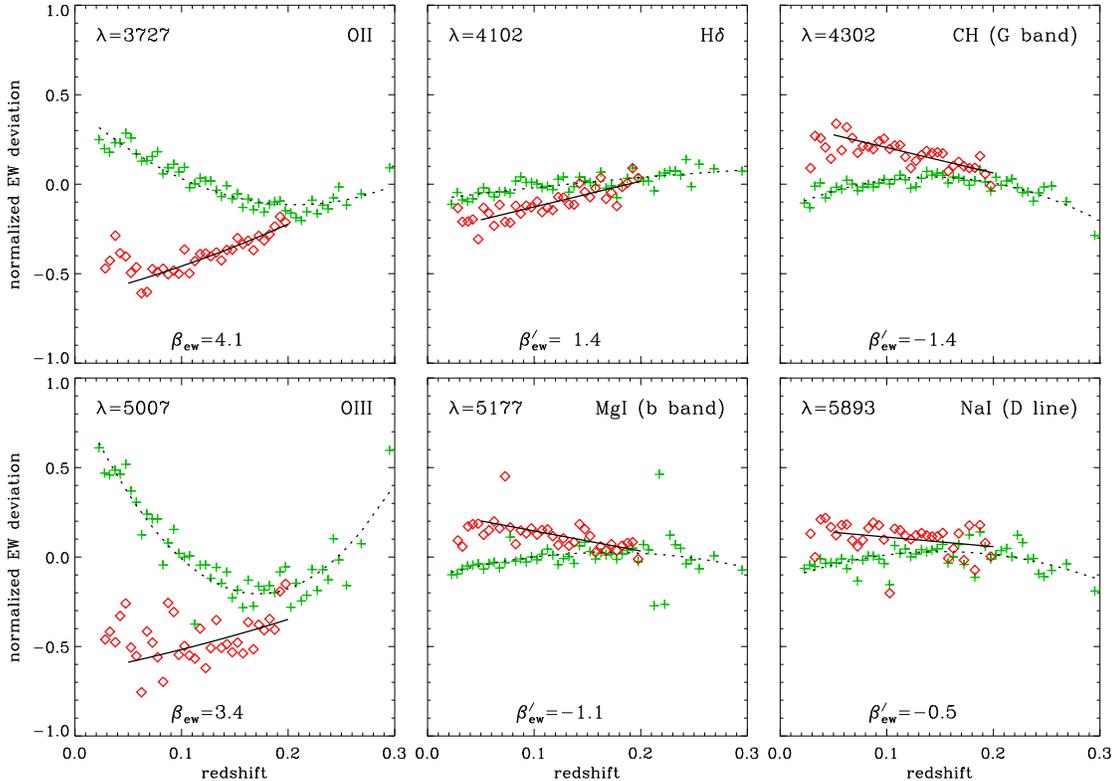}
}
\caption{Variation of average line equivalent widths{}
  -- O{\tiny II}, H$\delta$, CH, O{\tiny III}, Mg{\tiny I}, Na{\tiny
    I} -- as a function of redshift in the \grs\ data. The crosses
  represent 49 redshift bins between 0.020 and 0.32, selecting all
  galaxies. The dotted lines represent quadratic fits.  The diamonds
  represent a volume-limited sample of 35 redshift bins between 0.025
  and 0.20, selecting galaxies brighter than a k-corrected absolute
  magnitude of $-$21 (using $h=0.7$).  The solid lines represent
  straight-line fits over the redshift range where cosmic evolution is
  expected to dominate the measured variations.  The normalized
  deviation is defined as $(E - \langle E \rangle) / \langle E
  \rangle$ where $E$ is the equivalent width, and $\langle E \rangle$
  is the mean of the complete sample.  Note that the normalization is
  such that a positive value means an increase in emission for the
  emission lines (left panels) or an increase in absorption for the
  others.}
\label{fig:EW-lines}
\end{figure*}

How do the measured cosmic spectra vary as a function of redshift due
to selection effects, aperture effects, if any, and evolution?
Figure~\ref{fig:EW-lines} plots the equivalent width of O{\small
  II}, H$\delta$, CH, O{\small III}, Mg{\small I} and Na{\small I} as
a function of redshift for a complete sample and for a volume-limited
sample.  These measurements were made on luminosity-weighted, averaged
spectra determined from redshift bins of mostly 0.005 in extent.  To
interpret this figure, we first consider the volume-limited sample
(diamonds).
\begin{enumerate}
\item In the redshift range 0.05--0.2, we expect aperture effects to
  be minimal.  Therefore, the change in EW in this range could be
  primarily due to evolution and the changes agree qualitatively with
  a $\beta\sim3$ scenario.  It is not possible to make an accurate
  comparison due to the lower resolution of the model spectra.
  However, qualitatively they agree well: decrease in nebular-line
  emission between $z=0.2$ and 0.05; increase in CH absorption EW;
  less increase in Mg{\small I} and Na{\small I} absorption and, not
  surprisingly, the H$\delta$ absorption decreases due to a lower
  fractional contribution of A stars.  The solid lines in
  Figure~\ref{fig:EW-lines} are fits to these changes: with the ${\rm
    EW}\propto(1+z)^{\beta_{\rm ew}}$ for the emission lines, and
  ${\rm EW} = \beta'_{\rm ew} z + {\rm const.}$ for the others.  These
  constitute a detection of cosmic evolution. However, we cannot
  unambiguously separate cosmic SFR evolution from galaxy-population
  evolution and so we have not used these measurements to constrain
  the cosmic SFR in this paper.
\item In the redshift range 0.02--0.05, the change in EW for most of
  the lines levels off or reverses direction. As described above, we
  expect the evolutionary changes to continue from $z=0.1$ to 0.02.
  Our interpretation is is that at low redshift the fiber aperture
  will only see a small part of the galaxy which for a blue-targeted
  sample is biased towards more recent star formation -- thus, the
  nebular-line emission is higher and the CH absorption is lower.  We
  note that for an individual extended spiral galaxy, a 2dF-fibre
  spectrum could be biased towards more quiescent regions (e.g., the
  bulge) or more star-forming regions. Here, we are strictly refering
  to the average effect on the measured cosmic spectrum (the {\em
    median} bias may still be towards the bulge).  We note also that
  the nebular-line emission may be more sensitive to aperture and
  selection effects (and we ignore them in the fitting) than features
  at other wavelengths derived from older stellar populations.
\end{enumerate}

Considering the all-selection sample (crosses) from
Figure~\ref{fig:EW-lines}, it is clear that luminosity-selection bias
is having a more significant effect than aperture bias, i.e., the
difference between not including and including the lower luminosity
galaxies in the cosmic spectrum is larger than the difference between
using small and large physical apertures on a volume-limited sample
given the above interpretation.  This is most evident for the O{\small
  II} and O{\small III} emission lines showing that the less-luminous
galaxies contain a higher fraction of recent star formation.  The
\grs\ is a magnitude limited sample so the luminosity cutoff will rise
with redshift. This is illustrated in Figure~\ref{fig:sel-ap} -- at
$z=0.1$ we are probing 1.5 magnitudes fainter than the Schechter
luminosity function break $M^*$.  For a Schechter function with a
faint end slope of $-$1.19 \citep[the \grs\ value found by][]{madg02}
this will encompass about 70\% of the light at $z=0.1$, 95\% at
$z=0.04$ but only 20\% at $z=0.2$.

It should be noted that all magnitude-limited samples are dominated by
$M\sim M^*$ galaxies. This is a conspiracy between the volume probed
and the shape of the Schechter function \citep[e.g., see][for a
discussion of this]{glaze92}. Other studies of luminosity density
evolution via flux-limited samples will also be affected by this
dominance of $M^*$ galaxies, so the comparison of our results in
Section~\ref{sec:results-sub} is consistent.

If we consider our divided redshift ranges 0.03--0.11 and 0.11--0.25:
for the lower range, the luminosity selections are less significant
(the measured cosmic spectrum samples a significant range of the
galaxy luminosity function) but the aperture effects may be biasing
the averaged spectra towards more recent star formation; for the
higher range, the aperture effects are less significant but the
selection function is clearly missing a higher fraction of recent star
formation from the lower luminosity galaxies. However, in general, we
have found the results to be robust against choice of redshift range
chosen: lower, higher or combined.  In any case, the contours in
Figures~\ref{fig:ti-versus-ts}--\ref{fig:alpha-versus-r-2} take
account of the variation in best-fit parameters between the lower and
higher redshift ranges.

\section{Discussion and conclusions} \label{sec:conc}

We have developed a method of determining relative cosmic \sfh\ based
only upon the spectral information. We find consistent results between
both low-pass and high-pass spectral information and between physical
and empirical parameterizations of the star-formation law.
If we assume the following:
(a) the averaged \grs\ spectra ($z$ in the range 0.03--0.25)
represent the galaxy population as a whole; (b) the stellar
populations can be approximated by one chemical evolution scenario,
and; (c) the IMF, the PEGASE models and the cosmology are sufficiently
accurate; then we can reach the following conclusions.
\begin{enumerate}
\item The present day `cosmic spectrum'\footnote{It is interesting to
    compute the perceived color of the cosmic spectrum to the human
    eye, using standard color matching functions \citet{CIE71,CIE86}.
    Integrating these through the cosmic spectrum we have computed RGB
    values of 0.269, 0.388, 0.342.  This corresponds to a blue-green
    color, the closest match in standard RGB color lists, with the
    same color balance (normalisation or brightness is of course
    arbitrary), is `pale turquoise'. This is not a blackbody color,
    but this makes sense as the cosmic spectrum is a composite of
    young blue populations and old red populations.  It is robust
    against choice of redshift bins. See {\tt
      http://www.pha.jhu.edu/\~{}kgb/cosspec} for a rendering.} is well
  determined at $z\sim0.1$ and can only be fit with models
  incorporating consistent chemical evolution.  Constant metallicity
  is strongly ruled out. The final metallicity averaged on the
  luminosity is around solar ($Z\approx0.01$--0.04).
\item The star-formation timescale is longer than, or approximately
  equal to, the infall timescale of gas assembly.
\item The significant majority of nearby galaxies (weighted by their
  luminosity) have $\zform\ga0.65$, where $\zform$ is the redshift that
  star formation began.
\item There was a peak in cosmic \sfr\ density in the past, which was
  a factor of at least three times the present day rate.  With the
  physical parameterization, the peak occurs between $z=0.6$ and
  $z=10$ (using limits of $\zform \le 32$ and $\tin \ge 100$\,Myr).
  With the empirical parameterization, the peak occurs at $z=1$ or
  $z=5$ (with a instantaneous rise in the SFR at $z=5$).
\item Strong {\em upper limits} on star formation at high redshift
  ($\alpha$) can be obtained if we take inputs on the value of
  low-redshift star formation ($\beta$) from other studies
  \citep[e.g.][]{LLHC96}.  If we take $\beta\ga3$ (i.e., at least 8
  times the SFR at $z=1$ relative to $z=0$), then $\alpha<0.5$
  (Salpeter IMF, $\cal C$1).  These values are consistent with the no
  dust extinction SFR models of \citet{MPD98} and rule out the test
  case of a much larger SFR density at high redshift with a dust
  opacity that increases rapidly with redshift
  (their figure~7, which has $\beta\approx3$ and $\alpha\approx1$).
  The test case is also not consistent with our results for $\cal C$2
  which has $\alpha<-1$ for $\beta\ga3$. This upper limit corresponds
  to a maximum of 65\% and 55\% of the stars forming at $z>1$ in $\cal
  C$1 and $\cal C$2, respectively. Correspondingly, if we take
  $\beta\ga2$ then, a maximum of 80\% and 75\% formed at $z>1$.  Note
  that if there is any significant star formation prior to $z=5$, this
  will lower the value of $\alpha$ (or $\beta$) required to give a
  suitable model fit to the present-day cosmic spectrum.
\item Alternatively if we take the view that star formation was either
  constant or declining at high redshift (i.e. $\alpha\la0$) then we
  infer a rise in SFR to redshift unity, $1.5<\beta<5$, consistent with
  studies of luminosity density evolution.
\item From the 2MASS-\grs\ measurement of $\omstars$, \citet{cole01}
  concluded that their results were ``only consistent with recent
  determinations of the integrated cosmic star formation if the
  correction for dust extinction is modest.''  Here we find again that
  there can not be massive amounts of star formation concealed at high
  redshift by dust extinction.  \citeauthor{cole01}'s analysis
  compares the $J$ and $K_S$ luminosity density derived from \grs\ 
  redshifts and 2MASS photometry with semi-analytic models of
  galaxy/star-formation to derive an integrated stellar mass density
  today. Our method derives the star-formation history empirically
  using \grs\ spectra and redshifts.
\item From the \bj\ luminosity density of \citet{madg02} and computing
  the mass-to-light and SFR-to-light ratios from the PEGASE scenarios,
  we estimated the present-day stellar-mass density and SFR density.
  Restricting the models to good fits with $\beta\ga1.5$ and
  $\alpha\ga-3$, there is still significant degeneracy particularly
  with metallicity.  For cosmology $\cal C$1 with the
  \citeauthor{Salp55} IMF, $\omstars h$ is in the range 0.0020--0.0062
  and \rhosfr\ is in the range 0.024--0.065 \sfrunits, and with the
  \citeauthor{Kenn83} IMF, $\omstars h$ is in the range 0.0013--0.0033
  and \rhosfr\ is in the range 0.012--0.055 \sfrunits.  For cosmology
  $\cal C$2, the mass-to-light ratios increase and the values for
  $\omstars h$ are increased by a factor of about 1.2. The
  SFR-to-light ratios remain approximately the same. These results are
  consistent with the stellar-mass density derived by
  \citeauthor{cole01}\ and with other derivations of the local SFR
  density \citep[e.g.][]{sullivan00}.
\end{enumerate}

\begin{figure*}[ht]
\centerline{
\epsfxsize=9.2cm
\epsfbox{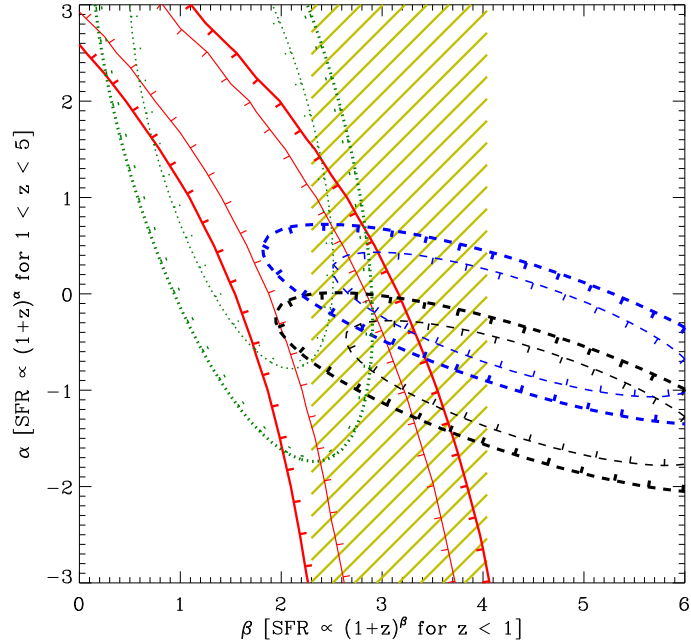}
}
\caption{Comparison between different cosmic SFR studies:{}
  confidence limits in $\alpha$ versus $\beta$. The {\em solid
    contours} represent the 2$\sigma$ and 3$\sigma$ confidence limits
  from this paper with FOM~B, $r=1.1$, Salpeter IMF and $C_0 =
  (0.70,0.3,0.7)$.  The {\em dashed contours} represent the limits
  from a compilation of UV luminosity-density measurements with a
  redshift range of 0.2--4.5 \citep{conn97,LLHC96,madau96,steidel99}, see
  \citeauthor{steidel99}'s figure~9. The lower and upper dashed contours
  are for ``no extinction'' and ``extinction corrected''. The {\em
    dotted contours} show the limits from a compilation of UV
  measurements with a $z$ range of 0.0--1.5
  \citep{CSB99,LPEM92,SLY97,treyer98}, see \citeauthor{CSB99}'s
  figure~13.  The {\em diagonally shaded region} represents the
  1$\sigma$ bootstrap confidence on $\beta$ from the compilation of
  various techniques by \citet{Hogg02} except we have excluded the UV
  measurements. The data includes nebular-line emission, far infrared
  and radio continuum measurements with a $z$ range of 0.0--1.0
  \citep{flores99,GZAR95,glaze99,HPWR00,hammer97,HCBP98,MCGH99,rowan97,TM98}.}
\label{fig:contours-all}
\end{figure*}

Finally, we compare our results with two different compilations of
cosmic SFR based on the rest-frame UV luminosity-density technique and
with a compilation of various other techniques. The $\alpha$-$\beta$
empirical parameterization was fitted to the data sets after converting
to cosmology $\cal C$1. The various contours are shown in
Figure~\ref{fig:contours-all} with full references quoted in the
caption.

For the first UV compilation, we took the 10 data points with error
bars from figure~9 of \citet{steidel99} and calculated formal confidence
boundaries in $\alpha$ versus $\beta$. The no extinction and
extinction corrected data were considered separately. This compilation
promotes a high value of $\beta\sim4$ with $\alpha$ depending on
extinction.

For the second UV compilation, we took 9 data points with error bars
from figure~13 of \citet*{CSB99} for the 2800\AA\ luminosity density
excluding those used in the first compilation. A minimum error of 0.07
in the log was used to calculate the boundaries. This compilation
promotes ``a gradual decline'' in the UV luminosity density for
$z\la1.5$ with $\alpha=\beta\sim1.5$.

For the other compilation, we recomputed the bootstrap error from the
measurements of $\beta$ given by \citet{Hogg02} excluding the UV
measurements included above. This new value was $\beta=3.2\pm0.9$.

A precise interpretation of Figure~\ref{fig:contours-all} depends on
the the accuracy of the errors used in the analysis.  Given the
analysis presented here, the two UV compilations (taking either
extinction model from the first) are consistent at the 3$\sigma$ level
with $1.8 < \beta < 2.9$ and $-1.0 < \alpha < 0.7$, but not quite
consistent at the 2$\sigma$ level. Our results are independently
consistent with these ranges of values but do not constrain the region
further.  Thus, evolutionary synthesis applied to the 2dFGRS-measured
cosmic spectra ($z\sim0.03$--0.25) is in concordance with the best-fit
results obtained from a variety of rest-frame UV luminosity-density
measurements ($z\sim0$--4.5).

\subsection{Future work}

With this technique of primarily using the `high-frequency' spectral
information, there are still significant degeneracies in determining
the cosmic \sfh.  However, combined with luminosity-density methods
and with further improvements, there is the potential to obtain more
accurate star-formation scenarios and to discriminate between IMFs and
cosmologies.  For example, the separation of galaxies into groups with
similar \sfh\ may help break the degeneracy, i.e., reducing the
merging of features between young and old populations. In addition, it
may be possible to improve the fit to the spectra by combining models
with different chemical evolution scenarios, e.g., weighted
combinations of spectra with different $\tin,\tsf,\zform$.

Future large surveys such as the SDSS main galaxy sample will provide
a consistency check with the \grs\ spectra (selected with different
effective wavelengths, 6200\AA\ versus 4700\AA, magnitude limits and
aperture diameters, see Figure~\ref{fig:sel-ap}).  With the \grs\ 
spectra (resolution 8--9\AA), and especially so with the SDSS spectra
(3--4\AA), it is apparent that higher spectral resolution
population synthesis models ($\sim$2\AA\ in the optical wavelength
range) are needed to maximize the scientific potential of these data
sets. Work is currently underway for this purpose.

Both galaxy surveys can also be used to measure the slope of
luminosity density variation with redshift in the local universe which
provides another star formation estimate from the same survey.  This
can done by either by using k-corrections and determining the
luminosity function in the rest frame or by determining the luminosity
function in observed wavelength, i.e., the extra-galactic background
light per unit redshift.  These measurements can be used to apply
further constraints on the cosmic star-formation scenarios and the
absolute cosmic comoving \sfr\ density.

\acknowledgements{The 2dF Galaxy Redshift Survey was made possible
  through the dedicated efforts of the staff of the Anglo-Australian
  Observatory, both in creating the 2dF instrument and in supporting
  the survey observations.  Karl Glazebrook and Ivan Baldry
  acknowledge generous funding from the David and Lucille Packard
  foundation.}


\end{document}